\documentclass[ aps,reprint, superscriptaddress,
                amsmath,amssymb,
                aps,
                prl,
                ]{revtex4-1}
\usepackage{graphicx}
\usepackage{dcolumn}
\usepackage{bm}
\usepackage{hyperref}
\hypersetup{
  colorlinks,
  linktoc=all,
  citecolor=green,
  linkcolor=red,
  urlcolor=blue}
\usepackage[mathlines]{lineno}
\usepackage[utf8]{inputenc}
\usepackage[table]{xcolor}
\usepackage[super]{nth}
\usepackage{rotating}
\newcommand\gr{\cellcolor{green!10}\hspace{3pt}}

\newcommand\ora{\cellcolor{magenta!10}\hspace{3pt}}
\newcommand\p{\cellcolor{orange!10}\hspace{3pt}}
\newcommand\y{\cellcolor{cyan!10}\hspace{3pt}}

\begin{document}


\title{Coupling an ensemble of electrons on superfluid helium to a superconducting circuit}

\author{Ge Yang} \affiliation{The James Franck Institute and Department of Physics, University of Chicago, Chicago, IL, USA}

\author{A. Fragner} \affiliation{Department of Physics and Applied
Physics, Yale University, New Haven, CT, USA}

\author{G. Koolstra} \affiliation{The James Franck Institute and Department of Physics, University of Chicago, Chicago, IL, USA}

\author{L. Ocola} \affiliation{Argonne National Laboratories, Center for Nanoscale Materials, Argonne, Illinois 60439, USA}

\author{D.A. Czaplewski} \affiliation{Argonne National Laboratories, Center for Nanoscale Materials, Argonne, Illinois 60439, USA}

\author{R.J. Schoelkopf} \affiliation{Department of Physics and Applied
Physics, Yale University, New Haven, CT, USA}

\author{D.I. Schuster} \email{David.Schuster@uchicago.edu}
\affiliation{The James Franck Institute and Department of Physics, University of Chicago, Chicago, IL, USA}

\date{\today}

\begin{abstract}
The quantized lateral motional states and the spin states of electrons trapped on the surface of superfluid helium have been proposed as basic building blocks of a scalable quantum computer. Circuit quantum electrodynamics (cQED) allows strong dipole coupling between electrons and a high-Q superconducting microwave resonator, enabling such sensitive detection and manipulation of electron degrees of freedom. Here we present the first realization of a hybrid circuit in which a large number of electrons are trapped on the surface of superfluid helium inside a coplanar waveguide resonator. The high finesse of the resonator allows us to observe large dispersive shifts that are many times the linewidth and make fast and sensitive measurements on the collective vibrational modes of the electron ensemble, as well as the superfluid helium film underneath. Furthermore, a large ensemble coupling is observed in the dispersive regime during experiment, and it shows excellent agreement with our numeric model. The coupling strength of the ensemble to the cavity is found to be $> 1$ MHz per electron, indicating the feasibility of achieving single electron strong coupling.

\end{abstract}

\pacs{Valid PACS appear here}
\maketitle

\section{Introduction}

Electrons on helium are a promising resource for quantum optics and quantum computing\cite{Lyon2006,Platzman1999,Dykman2003,Schuster2010b}. They form an extremely clean two dimensional electron gas\cite{Monarkha2004}, as evidenced by a mobility exceeding \(10^7 \,\mathrm{cm^2/V s}\) \cite{Shirahama1995a,Shirahama1995b}. and the electron spin coherence time is predicted to exceed \(10^3 \,\mathrm s\) \cite{Lyon2006}. Electrons on helium have been used to study Wigner crystallization and quantum melting \cite{williams1971, Grimes1979,Andrei1997}. Recent experiments employ them as a powerful probe to study the topological domain structures on the surface of superfluid helium 3 \cite{dwyer2013, kono2010, chepelianskii2015}. In addition, it is now possible to build mesoscopic structures such as quantum dots with one or a few electrons on helium, and single electron scale charge coupled devices \cite{Papageorgiou2005, Rousseau2009, Bradbury2011, Takita2014}. However, performing quantum experiments in this fascinating system has lagged behind that in semiconducting 2D electron gasses, such as GaAs, as traditional measurement techniques cannot be applied to electrons on helium. In particular, it is not possible to make direct Ohmic contact to the electron gas.  Additionally, the largely unscreened electron-electron Coulomb force and a hydrostatic instability of the system \cite{Gorkov1973,Edelman1980} suppress the exchange interactions typically used in semiconductor spin qubits \cite{Petta2005a}.

The circuit QED architecture \cite{Wallraff2004, Blais2004} offers a path to new experiments in the quantum regime as well as improving the sensitivity and bandwidth of existing measurements. In this hybrid approach, electrons are trapped above an on-chip superconducting microwave resonator. The presence of the electrons changes the effective capacitance of the cavity, resulting in a dispersive shift of the cavity resonance frequency. In the strong dispersive regime, the cavity frequency shift is larger than the cavity linewidth, and every photon measures the state of the electrons. Because the energy of a single photon in the cavity is higher than the thermal bath (\(\hbar \omega > k_b T\)), it is possible to conduct quantum optics experiments at the single photon level. This dispersive measurement is conceptually similar to the Sommer-Tanner technique \cite{Sommer1971}, but the use of resonant superconducting circuits at microwave frequencies enables better impedance matching, resulting in faster and more sensitive measurements of small ensembles. Finally, the hybrid architecture allows one to leverage the substantial progress in superconducting circuits over the past decade \cite{Schoelkopf2008, Devoret2013}.

In this letter, we report the first implementation of a circuit QED architecture with electrons on helium. We show lithographic control and sub-nanometer measurement of the superfluid helium film thickness.  Our experiment shows a strong dispersive shift due to the electrons that is many times the cavity linewidth. On average, the coupling per electron in the ensemble is about 1 MHz, suggesting single electron strong coupling should be within reach. Electrons can be held for many hours, and their normal mode frequencies and number can be controlled by adjusting the trapping potential. The resulting evolution of the dispersive shift agrees excellently with our numerical model.

\section{Experimental Setup and detection technique}
\begin{figure}[b]
\includegraphics[width=3.3in]{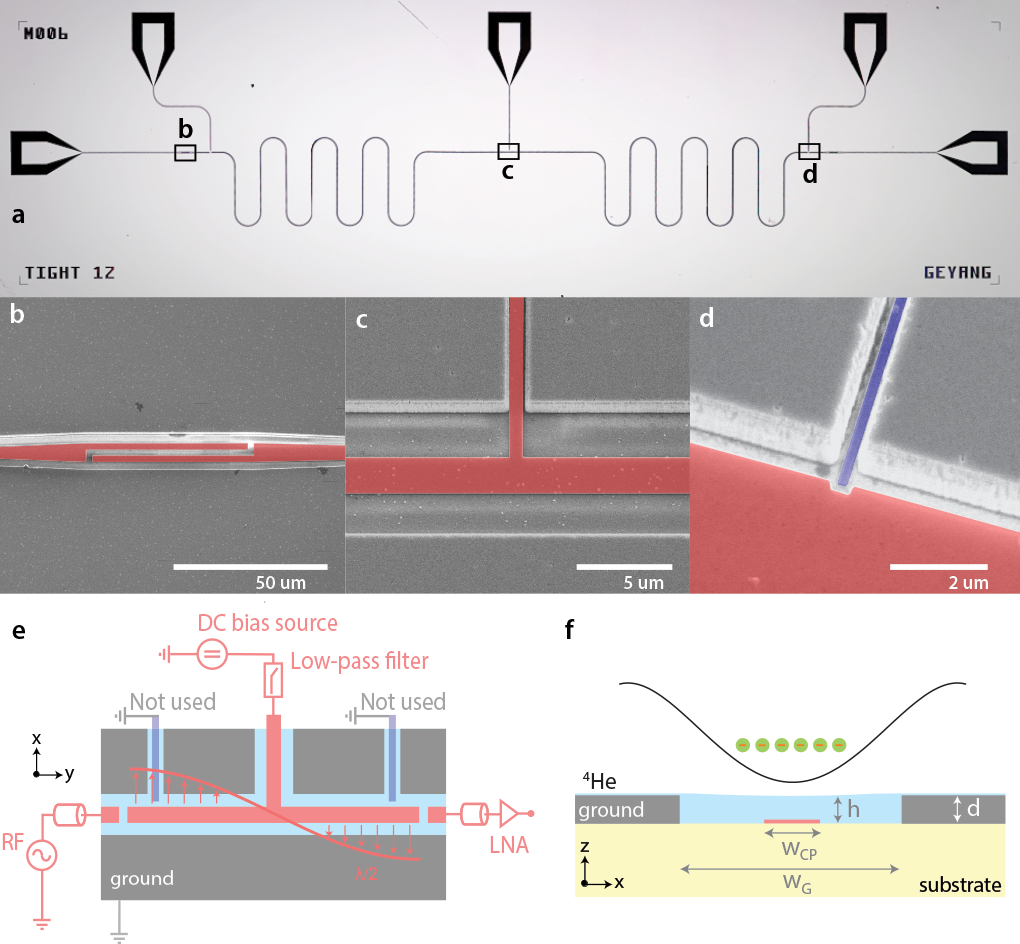}
\caption{Device, circuit schematic and trap geometry. \textbf{a}) Optical and SEM images of a cavity-electron ensemble trap on a $2\times 7$mm superconducting chip. The device is positioned 5.5 mm above the bottom of a cylindrical superfluid reservoir of radius $r=3.175$ mm, mounted in a hermetically-sealed copper box at 25 mK in a dilution refrigerator. \textbf{b}) Interdigitated gap capacitors with gap width 2 $\mu$m at the cavity input. \textbf{c}) DC bias electrode connected directly to the center pin of the cavity at a node of the standing wave voltage distribution of the fundamental mode. \textbf{d}) Sub-$\mu$m-size electron trap near voltage maximum of the fundamental mode with constriction of width 500nm. \textbf{e}) Circuit schematic showing the voltage distribution of the fundamental mode (red) and the simplified measurement and control circuit connected to the  center pin (pink). The cavity is measured in transmission using a low-noise amplifier and the trap potential is tuned through a DC source connected to the center pin (pink) through a low-pass filter. \textbf{f}) Cross-sectional view of the cavity waveguide gap showing the schematic trap geometry. The ground planes (gray) form a micro-channel of height $d=800$ nm and width $w_{\rm G} = 6\,\mu$m filled with superfluid $^4$He by capillary action. A DC voltage on the submerged center pin (pink) of width $w_{\rm CP} = 2\,\mu$m and thickness $t= 80$ nm creates a parabolic trapping potential for electrons above the surface which couple to the RF field in the cavity. \label{fig:ResonatorImage} } 
\end{figure}

The electron on helium circuit QED setup consists of an integrated electron trap and coplanar waveguide (CPW) resonator (see Fig.~\ref{fig:ResonatorImage}a). The ground planes of the resonator are thicker than the center pin, forming a micro-capillary channel which determines and stabilizes the superfluid helium film thickness  \cite{Marty1986,Rees2011}.  The electrons are held in the resonator volume by a DC bias voltage applied to the center pin as shown in Fig.~\ref{fig:ResonatorImage}c.

The electrons are confined in both the transverse and longitudinal direction of the channel. In the transverse direction, a DC-voltage applied to the center pin creates a parabolic
trapping potential (Fig.~\ref{fig:ResonatorImage}c) that confines
the electron ensemble in the channel, colocated with the microwave
field. 
In addition to the large electron trap formed by the resonator center pin,
the devices also contain smaller $\mu$m-size electron traps positioned
near the voltage maxima of the fundamental mode for future
single-electron experiments (Fig.~\ref{fig:ResonatorImage}d). Those
smaller traps were set to ground potential throughout the experiments
discussed in the rest of this paper. The input and output of the resonator (Fig.~\ref{fig:ResonatorImage}b) side of the coupler are held at 0 V to prevent electrons from leaking out the sides.  Along the cavity DC bias lead, where the potential may be positive, constrictions shield the electrons ensuring that there is a potential barrier for escape.

The bare cavity resonance frequency is $\omega_0/2\pi \simeq 
4.789$ GHz, loaded quality factor $Q_L\simeq 17750$ and corresponding decay 
rate $\kappa/2\pi\simeq 270$ kHz in the absence of any superfluid or
electrons. The Q of the sample is set by the
couplers, not by the internal Q of the resonator, despite the fact that the DC bias lead directly connects the center pin to a low impedance. This is possible because the connection is made at a voltage node, where radiation is minimized \cite{Armour2013, Petersson2012}

\begin{figure}[h] \includegraphics[width=3.4in]{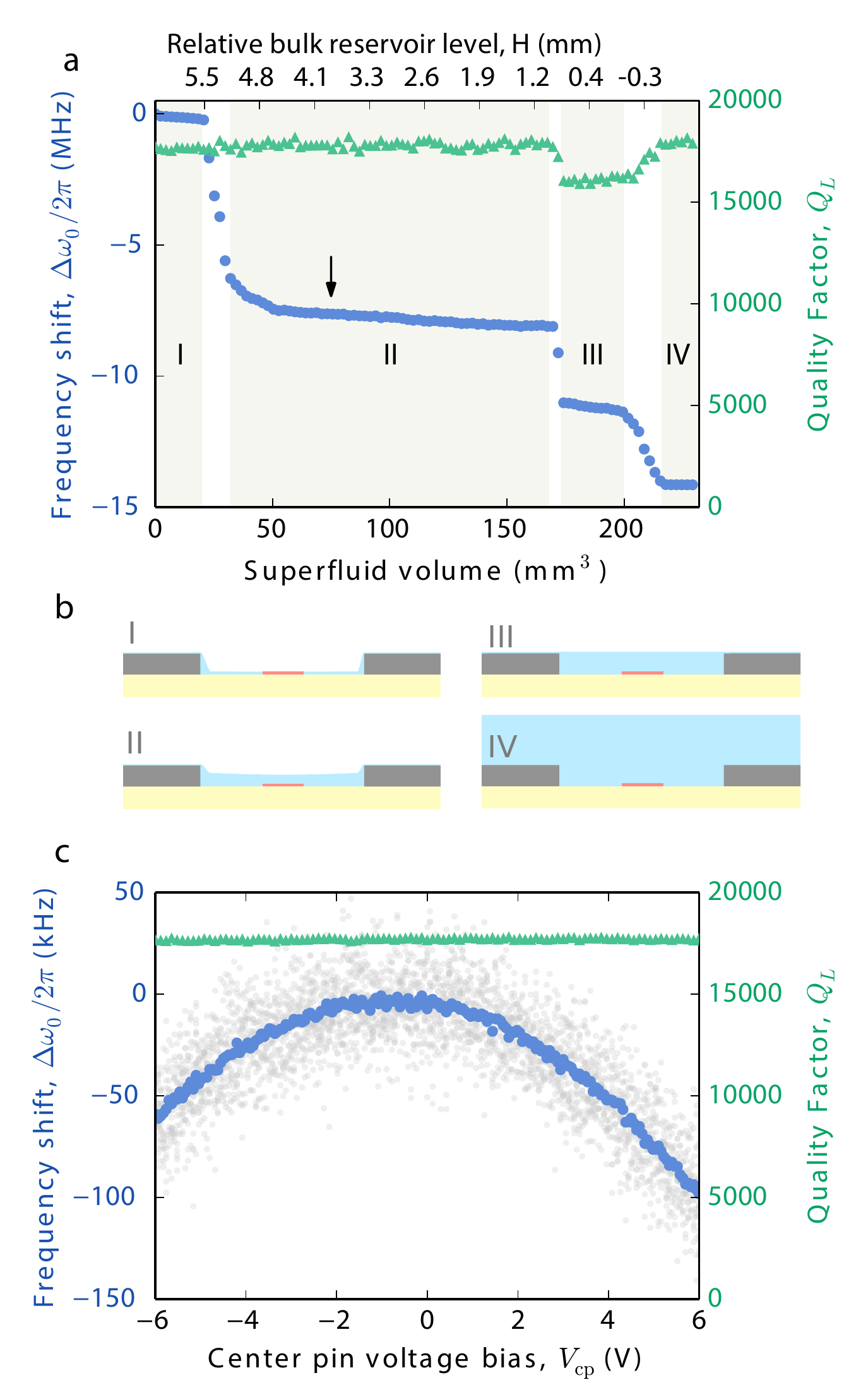}
\caption{\label{fig:LevelMeter} Cavity response to superfluid helium.
\textbf{a}, Measured resonance frequency shifts $\Delta\omega_0/2\pi$
(blue dots, left axis) and loaded quality factor $Q_L$ (green triangles,
right axis) as functions of superfluid volume supplied to the cell
(bottom axis) and relative bulk helium level $H$ in the reservoir (top
axis). Each datapoint corresponds to an increase in superfluid volume of
$\Delta V_{\rm sf} \sim 2.3$ mm$^3$ and reservoir level $\Delta H\sim
70\,\mu$m. \textbf{b}, different filling state corresponding to the different
regimes in \textbf{a}. \textbf{c}, Frequency shift (blue dots, left axis) and
quality factor (green triangles, right axis) as functions of center pin
voltage bias $V_{\rm cp}$ at fixed helium level in the capillary action
regime indicated by arrow in \textbf{a}.  Gray dots are frequency shifts extracted from single-shot cavity
transmission measurements with $N=80$ such measurements per voltage bias
point $V_{\rm cp}$. The blue data points are averages over the
single-shot measurements at each point.} 
\end{figure}

\section{Helium Dynamics}

An important prerequisite for trapping electrons on helium in a 
micro-channel geometry is to establish a self-stabilized film of superfluid 
helium of known thickness, which can be achieved by capillary action filling
of the channels from the low-lying bulk reservoir. The helium raises the 
effective dielectric constant of the waveguide, lowering the resonator 
frequency proportionally to the thickness, $h$.

To measure the cavity response to superfluid helium, we monitor the
resonance frequency and quality factor in transmission while increasing
the bulk helium reservoir level in small increments (the center pin is
held at ground potential throughout this measurement). The results of
such a helium filling experiment are presented in
Fig.~\ref{fig:LevelMeter}\,a.  Four different regimes can be clearly distinguished in the frequency shift. For small amounts of superfluid (regime I), an unsaturated van-der-Waals (vdW) film of
thickness $h\sim 30$ nm forms as the liquid evenly coats the surface of
the resonator and the interior of the sample cell, leading to small
frequency shifts of $\Delta\omega_0(h)/2\pi\simeq -190$ kHz. Once the vdW film has saturated, the liquid film shape is determined by capillary action with a semi-circular profile $z(x)\sim x^2/2R_c(H)$ in
the gap. The capillary radius $R_c(H)=\sigma/\rho gH$ is determined by the 
distance between the bulk helium level in the reservoir and the chip surface
$H$, where $\sigma=0.378\times 10^{-3}$ N/m is the surface tension of liquid
helium in vacuum, $\rho=0.154\times 10^{-3}$ kg/cm$^3$ the mass density and 
$g$ the gravitational acceleration. When the
radius of curvature becomes on the order of the gap width $R_c \sim
w_G$, the gap starts to fill up by capillary action and is filled
completely for $R_c \gg w_G$ (regime II).  Small increases
in shift in the subsequent ``flat" regime are due to decrease in the curvature of the helium profile. Finite element simulations show that when the channel is filled ($h=800$ nm) the frequency shift is $-8$ MHz, in good agreement with the observed data. As $H\rightarrow 0$, the radius of curvature becomes on the order of the chip dimensions and eventually starts to diverge.  We attribute the abrupt jump at 170 mm$^3$ to the formation of a thick film which spans the entire chip (regime III) and is supported by the sample holder. As $R_c\rightarrow\infty$, the superfluid film becomes sensitive to mechanical vibrations and small fluctuations in the reservoir level which manifests itself in a perceived drop in quality factor in this regime. Once the reservoir has been completely filled, the helium fills the region above the chip linearly (regime IV) until the resonator becomes insensitive at thickness $h\sim 6\,\mu$m, corresponding to a frequency shift of $\Delta\omega_0(h)/2\pi=-14.145$ MHz, again in good agreement with numerical simulations that predict a final frequency shift of \(-14.1\) MHz. All subsequent experiments are done at the filling level indicated by the black arrow in Fig.\ref{fig:LevelMeter}\,a, which corresponds to \(h \approx 647\) nm where the frequency shift is  
$\Delta\omega_0(h)/2\pi\simeq -7.4$ MHz, and is only slightly changed by additional fluid introduced to the reservoir.  

The superfluid level in the resonator gap can be modulated in-situ by
sweeping the voltage of the center pin $V_{\rm cp}$. To lowest order,
the equilibrium film thickness at the center of the gap is determined by
the electromechanical force on the film surface and surface tension with
a quadratic voltage dependence $h(V_{\rm cp}) \approx h(0)+ (V_{\rm
cp}^2/16\sigma)(\varepsilon_{\rm He}-\varepsilon_0)$.
Fig.~\ref{fig:LevelMeter}\,b shows measured frequency shift and quality
factor as functions of center pin voltage at a reservoir level of
$H\simeq 4$ mm where the gap is partly filled by capillary action. The
resonance frequency shows a parabolic voltage dependence while the
quality factor remains constant, as expected. The maximum observed shift
at $V_{\rm cp} =+6$ V of $\Delta\omega_0/2\pi = -100$ kHz corresponds to
a change in film thickness of $\Delta h \simeq 13$ nm at the center of
the gap.  The slight offset of the other-wise symmetric response
is not well-understood, and it is device dependent. The frequency sensitivity to level changes in the gap can be estimated from finite element electromagnetic simulations to be $ \approx 8$ kHz/nm, consistent with the overall slope and frequency shift.
The stability of the capillary action film is
estimated from consecutive single-shot transmission measurements (gray
data points in Fig.~\ref{fig:LevelMeter}\,b), with $N=80$ frequency
measurements per voltage bias point. Slow fluctuations of the helium level are manifested in the $\delta\omega_0^{\rm (rms)}/2\pi = 16$ kHz scatter of resonance frequencies (gray points in Fig.~\ref{fig:LevelMeter}b)   
corresponding to helium level fluctuations of $\delta h^{\rm
(rms)}\approx 2.6$ nm. 

In summary, the microwave measurement provides a high bandwidth way to measure the helium level and its fluctuations, down to ${\rm pm}/\sqrt{\rm Hz}$ level sensitivity. Using this technique, we establish a lithographically defined, stabilized superfluid helium film within the cavity-trap. In the regime of the experiment, the channel helium level is insensitive to the small differences in the amount of helium put into the sample box. The measured helium level fluctuations are relatively small and will be monitored to see if they have a significant effect on the electron coherence time through changes in the trapping potential.

\begin{figure}[th] \includegraphics[width=3.4in]{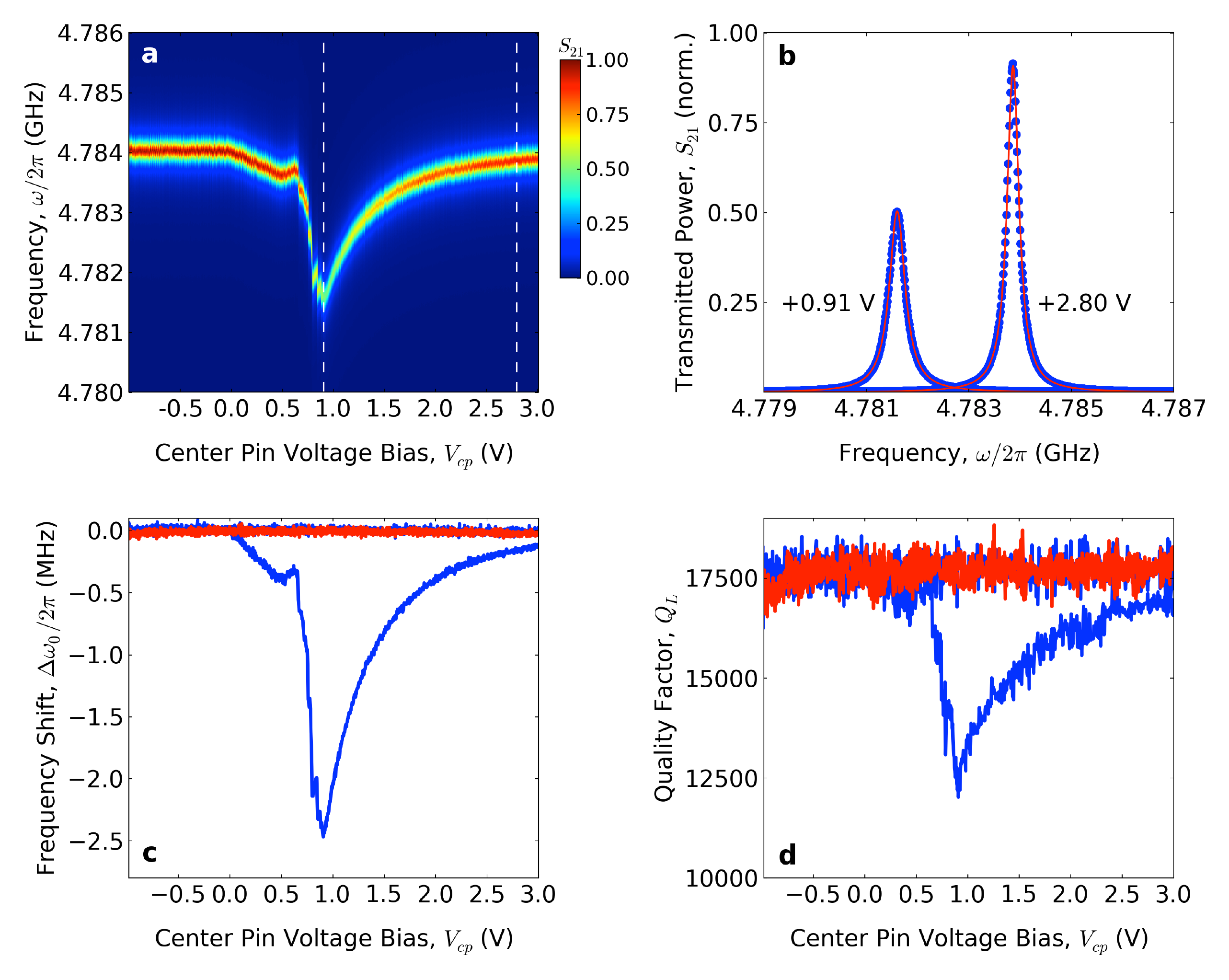}
\caption{Detection of a trapped electron ensemble on superfluid helium
in a cavity transmission experiment. \textbf{a}, Normalized transmitted
power through the cavity as a function of center pin trap voltage $V_{\rm
cp}$. \textbf{b}, Normalized transmission spectra at $V_{\rm cp}=+0.91$
V and $+2.8$ V, showing a shift in resonance frequency and a reduction in
transmitted power at the bias points indicated by the dashed vertical lines
in \textbf{a}. Solid red lines are fits to Lorentzians. \textbf{c,d},
Resonance frequency (\textbf{c}) and loaded quality factor (\textbf{d})
as functions of trap bias in the presence (blue) and absence (red) of an
electron ensemble. In Figs.~\textbf{a} - \textbf{d}, electrons are first
loaded into the cavity mode volume at an initial bias of $V_{\rm
cp}=+3$V and a fixed helium level in the capillary regime with an
uncharged shift of $\Delta\omega_0/2\pi = -7.58$ MHz and a reservoir
level of $H\simeq 4$ mm. The blue line shows the trap voltage being
swept from +3 to -1V and back in 4 mV steps, eventually depleting the
trap region, while the red line shows the same sweep for an empty trap.}
\label{fig:ElectronDetection} \end{figure}

\section{Dispersive measurements of electrons in a cavity}
Following uncharged superfluid measurements, we proceed to load electrons into the cavity mode volume and detect the trapped electron ensemble in transmission measurements. Electrons are generated via pulsed thermionic emission from a tungsten filament mounted in vacuum above the device and attracted towards the superfluid surface in the resonator channel by a positive trap voltage $V_{\rm cp}$. After waiting for the sample to cool, the cavity transmission is monitored while tuning the center pin voltage $V_{\rm cp}$ (Fig.~\ref{fig:ElectronDetection}) starting from $+3$V. The dispersive interaction of the cavity with the trapped ensemble leads to a voltage-dependent shift of the cavity resonance towards lower frequencies before reverting back at negative trap potentials. We observe maximum resonance shifts of up to $\Delta\omega > 10 \kappa$ cavity linewidths in frequency while Q is somewhat reduced (Fig.~\ref{fig:ElectronDetection} b). The electron-induced frequency shift reaches a maximum of $\Delta\omega_{\rm max}/2\pi=-2.47$ MHz at $V^{\rm (th)}_{\rm cp}=+0.91$ V with a drop in quality factor and a corresponding increase in cavity decay rate of $\Delta\kappa_{\rm max}/2\pi = 122$ kHz (blue curves in Fig.~\ref{fig:ElectronDetection}\,c and d). These changes in the cavity resonance frequency are at least an order of  magnitude larger than those caused by the electric field induced helium film thickness change without electrons. Below the threshold $V^{\rm (th)}_{\rm cp}$, the electron-induced shifts decrease gradually as electrons are lost from the trapping region. To ensure that the observed cavity response is due to the trapped electron ensemble, we performed a control experiment where the filament was fired while the center pin was biased at -1V. The voltage is then swept in the reverse direction (red curves in Fig.~\ref{fig:ElectronDetection}\,c and d). The voltage dependent signal is completely absent (see Fig.~\ref{fig:LevelMeter}\,b and discussion above). The ensemble-induced cavity response has been reproduced in independent experiments using five different devices. The maximum observed resonance shifts are repeatable and generally vary between $2 - 8$ MHz based on loading conditions.

\begin{figure}[th]
  \includegraphics[width=3.4in]{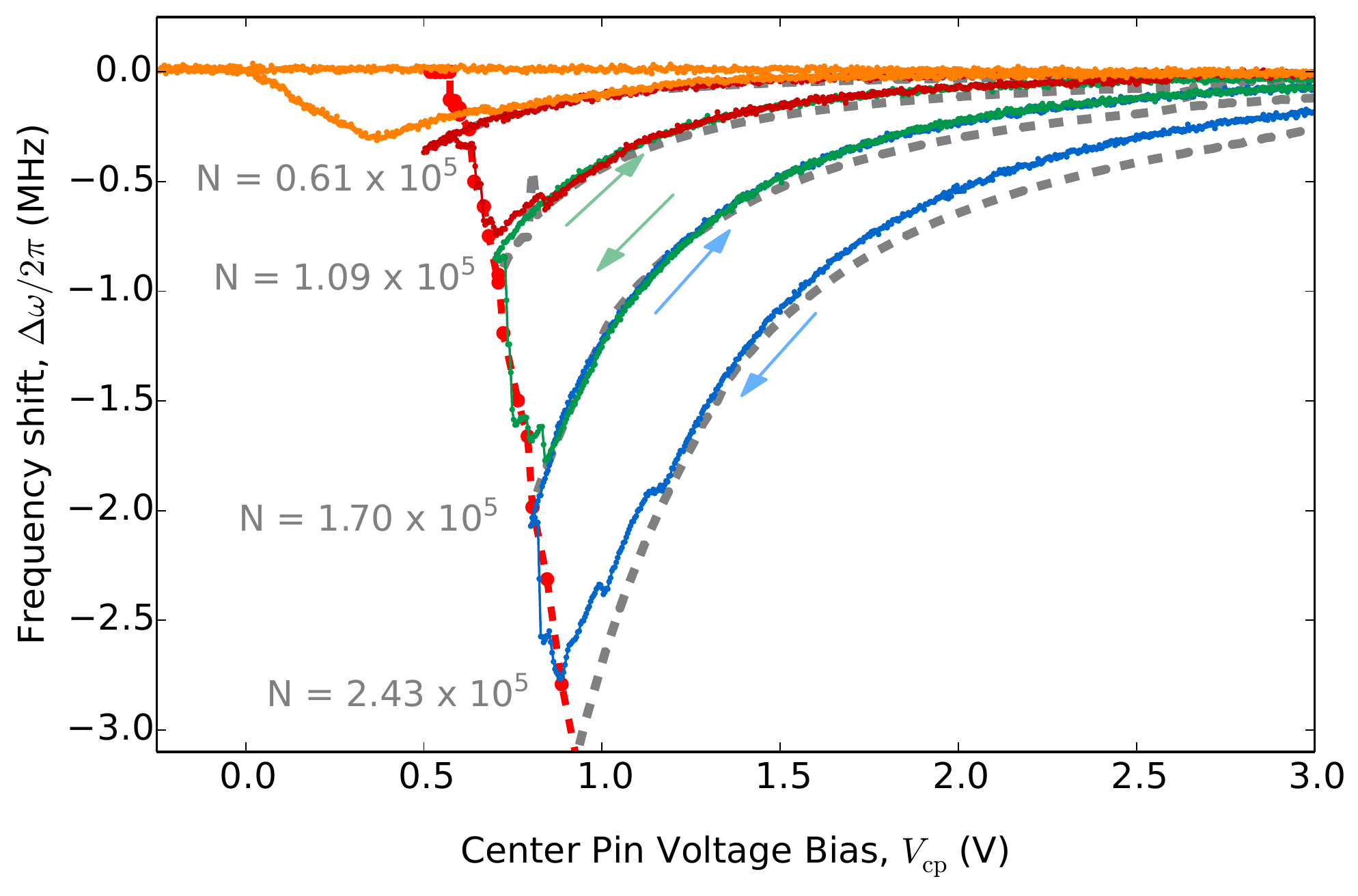}
  \caption{Measured cavity frequency shifts as a function of trap voltage bias $V_{\rm cp}$. Colors indicate consecutive cycles of the voltage sweep. In each cycle $V_{\rm cp}$ is decreased (arrows) until electrons are irreversibly lost from the trap and then increased to the initial value of 3.0 V. Gray dashed lines in the background show frequency shifts predicted by molecular dynamics simulation. For each of the four iso-electron number curves the number of electrons in the simulation is depicted to the left. A red dotted line indicates modeled electron loss with respect to a leak voltage of \(530\) mV. The iso-electron number curves from the experiment terminate within a small neighborhood of this loss frontier.} \label{fig:HysteresisLifetime} 
\end{figure}

To further investigate the quantitative form of the cavity shift  as a response to the number of electrons \(N\) and the bias voltage  \(V_\text{cp}\), we designed a protocol to partially drain the electrons from the trap. In the following experiment at the end of the first sweep from +3V (bottom half of blue curve in Fig.~\ref{fig:HysteresisLifetime}), we deliberately sweep down below the \(V_\text{cp}^\text{(th)}\) to introduce electron loss. Now with fewer electrons in the trap (top half of blue curve in Fig.~\ref{fig:HysteresisLifetime}), the cavity resonance shift is smaller in magnitude than before, but non-hysteretic unless another electron-loss event is triggered. For subsequent sweeps, we then set successively lower stop voltages. The final sweep (orange curve in Fig.~\ref{fig:HysteresisLifetime}) depletes the mode volume of all electrons as evidenced by the vanishing frequency shift on the final upward sweep. By carefully tuning the stop voltages and number of sweeps we can controllably reduce the number of electrons as desired.


A quantitative understanding of the electron-cavity interaction requires modeling of the classical many-body interactions between electrons as well as the coupling of the resulting electron normal modes with the cavity. We develop a non-perturbative numerical model which determines the electron ensemble configuration, frequencies and coupling to the cavity. First simulated annealing is employed to find the minimal energy configuration of the electrons. The electrostatic potential used in this step is constructed using field profile data derived from the sample geometry. After obtaining the electron configurations, we solve the equations of motion of the cavity-electron coupled system in a non-perturbative way to obtain the cavity frequency shift. For a given number of electrons in the trap, this calculation is repeated for various \(V_\text{cp}\) to produce the corresponding frequency shift curve. Curves for various number of electrons are computed, and no fitting parameters are used in the model besides picking the closest iso-electron number curve. Fig.~\ref{fig:HysteresisLifetime} shows excellent agreement between the data and our computational model. The model is described in detail in the supplementary material.

Using the measured signal and our model, we can infer the electron mode frequency and coupling strength $g$. The frequency of the strongest coupled mode grows proportionally to \(\sqrt{V_\text{cp}}\) and is roughly 25 GHz at $V_\text{cp} = 0.5$ V, which shows we are well within the dispersive regime.  In this regime the coupling is linearly proportional to the distance of the electron to center of the trap, approximately \(1.6\,\text{MHz/}\mu\text{m}\) per electron. In future trap designs, additional guard electrodes can improve trap stability at low voltage allowing access to the resonant regime. The coupling though quite large already, can be enhanced further by shrinking the dimensions of the trap.

There are two types of electron loss observed. The first, which determines the number of electrons loaded, occurs at higher $V_\text{cp}$ due to hydrodynamic instability\cite{Marty1986}. The density immediately after loading at $V_\text{cp}=3$V was $n\approx 2\times 10^9\, \text{cm}^{-2}$. The second type of electron loss occurs when the trap depth becomes sufficiently shallow, such that electrons can leak out of the trap. We model this phenomenologically by assuming that electrons are lost if the potential difference between ground plane and electrons is less than $V_{\text{leak}}$. Using $V_{\text{leak}}$ as a single fit parameter in the molecular dynamics simulation, we find best agreement between simulation and experiment when $V_{\text{leak}} = 530$ mV. The residual population for small electron numbers at small $V_{\text{cp}}$ (orange curve in Fig.~\ref{fig:HysteresisLifetime}) is not well understood. 

In summary, we have demonstrated the successful trapping and detection of an electron ensemble above the surface of superfluid helium in the circuit QED architecture. The measurement technique introduced here could extend traditional electrons on helium experiments to smaller ensembles and enable observation of the electron dynamics. The observation of the large dispersive shift and the good agreement with our numerical simulations indicate that it should be possible to perform cavity QED experiments in a single electron quantum dot. Though small, the fluctuations in the helium film thickness are an important source of decoherence for the electron motional states, and merit further study. Finally, the sensitivity of the device to helium thickness changes can be exploited for novel cavity optomechanics experiments with superfluid ripplons.

\begin{acknowledgments} The authors thank Mark Dykman, Steve Lyon, David G. Rees for many helpful discussions, and the HOOMD molecular dynamics simulation package \cite{Anderson2008, HOOMD}. This work was supported by NSF CAREER grant DMR 1151839, the University of Chicago MRSEC program of the NSF under Award
No. DMR 1420709, and the David and Lucile Packard Foundation. 
Use of the Center for Nanoscale Materials, an Office of Science user facility, was supported by the U. S. Department of Energy, Office of Science, Office of Basic Energy Sciences, under Contract No. DE-AC02-06CH11357.
\end{acknowledgments}

\noindent\rule{8cm}{0.4pt}

\newpage
\onecolumngrid
\setcounter{figure}{0}    
\setcounter{equation}{0}    
\setcounter{table}{0}   
\renewcommand{\theequation}{S\arabic{equation}}
\renewcommand{\thetable}{S\arabic{table}}
\renewcommand{\thefigure}{S\arabic{figure}}
\appendix
\section{Supplemental information}

The situation where many interacting electrons in an electrostatic trap couple to a microwave resonator represents an interesting physical system. Experimentally it is not possible to directly observe how electrons arrange themselves inside the trap. However, such configurations contain important information of the electrostatic and microwave properties of the system. Therefore, the goal of this supplement is to develop a set of tools that provides a complete quantitative understanding of the classical interaction between few electrons in a trap and a microwave resonator. The first step is to start with an electrostatic simulation and solve for the equilibrium position of each individual electron in a given trapping potential.

With the electron configuration at hand, we can then calculate the dispersive shift of the cavity. We take a  non-perturbative approach to directly compute the cavity frequency from the equation of motion of the electrons-cavity coupled system. The numerical results are confirmed by a simple analytic model in the low density regime. In the high density regime where our experiment lies, the simulation model agrees well with the experimental data. Using this model, we are able to derive a number of relevant quantities, such as the total number of electrons on the resonator, and the average coupling per electron.

The fabrication recipe and experimental setup can be found at the end of the document.

\section{Electrostatic Simulation}
The first step of our modeling effort starts with the electrostatic simulation of the trapping potential. We simulate the trapping potential for a CPW in Maxwell, a finite element simulation tool. Since the geometry does not change in the $y$-direction, we only extract the cross-sectional profile. This numerical potential profile is then fitted with a \nth{14} order polynomial (with only the even power terms). From this we extract the second order coefficient to construct an ideal parabolic potential for use in the molecular dynamics simulation. Our analysis shows that for the most part, the electrons stay within 500 nm from the center, where this approximation is good.

\begin{figure}[h!]
\centering
\includegraphics[width=.8\textwidth]{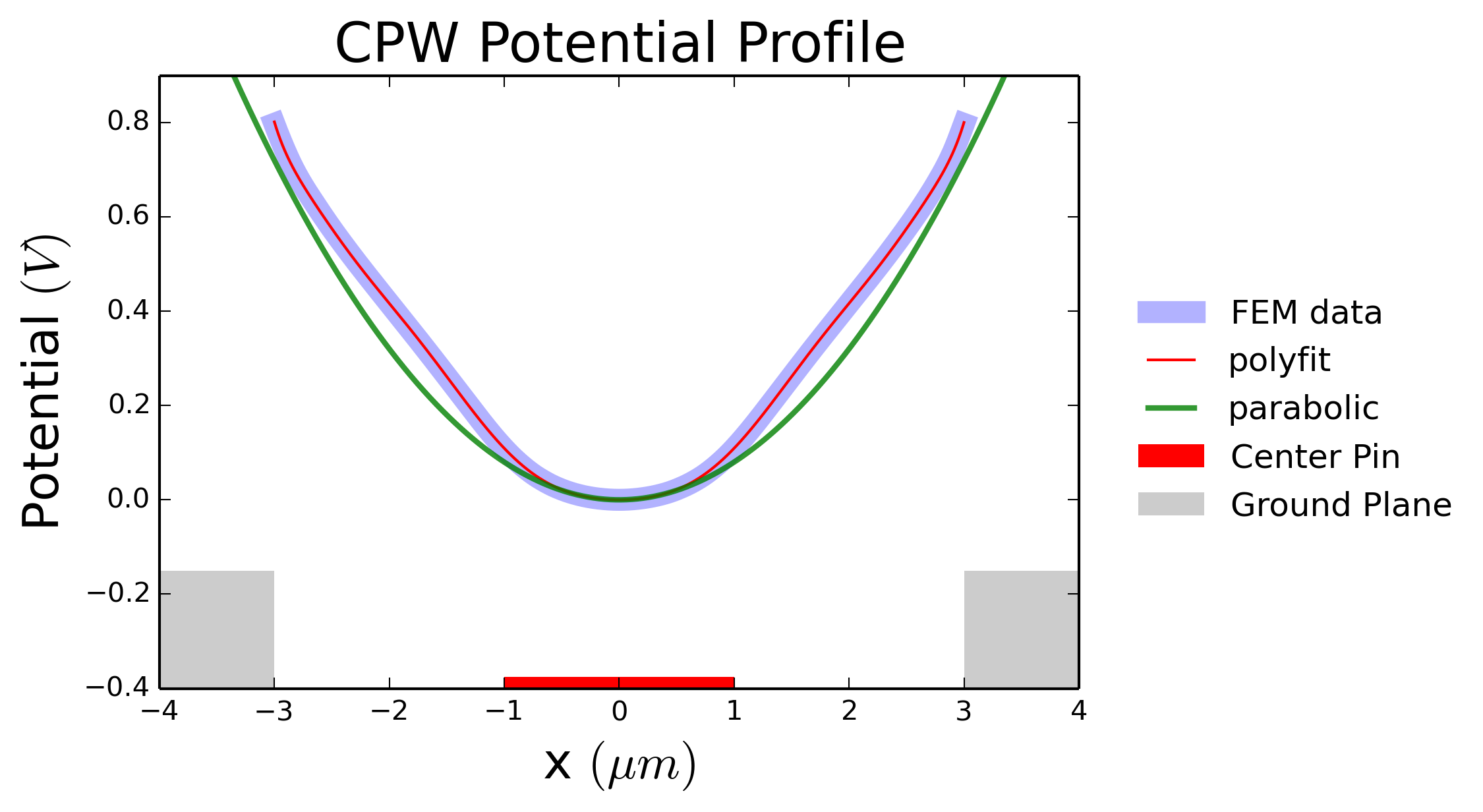}
\caption{Potential profile of the trap, showing the original numerical profile derived from a finite element simulation model; a polynomial fit,and a parabolic approximation using the second order term of the polynomial fit.}
\end{figure}

Numerically, the electrostatic potential used in the molecular dynamics simulation is
\begin{align}
    U(x, V_{\text{cp}}) = 0.0733 \,e V_{\text{cp}}  x^2, \label{eq:static_potential_function}
\end{align}
where $x$ is measured in $\mu$m from the center of the trap and $V_\text{cp}$ is the bias voltage applied to the center pin.

\section{Molecular Dynamics}

To solve for the equilibrium positions of the electrons in a given electrostatic potential, we use a simulation method called molecular dynamics. Molecular dynamics represents a class of deterministic, iterative algorithms
that can be used to find solutions to many-body problems. For this work, we use the package HOOMD \cite{Anderson2008}. This \textit{general-purpose} particle simulation toolkit scales on multi-core CPU and GPUs which allows us to quickly anneal up to 4000 electrons.

In a typical simulation run, we start with a fixed number of electrons $n$. Due to the large number of pairwise interactions that grows quadratically with $n$, we shrink our 12 mm long resonator from the actual length down to a \(50\,\mathrm{\mu m} \times 50 \,\mathrm{\mu m}\) box with periodic boundary conditions on each side. Simulation and experimental results can then be compared by multiplying the number of electrons in the \(50\,\mathrm{\mu m} \times 50 \,\mathrm{\mu m}\) box by $L_{\text{res}}/L_\text{box} \approx 243$. In the remainder of this supplement we refer to the number of electrons on the resonator as $N=243 \cdot n$. 

There is no long-range screening in our simulation. However, the Coulomb interaction is cut off at \(20 \,\mathrm{\mu m}\) to prevent electrons from interacting with their own image charges across the periodic boundary condition, causing an explosion of pairwise interactions in the system.

The lowest energy electron configuration is found by annealing the system. Here the temperature of the ensemble is gradually decreased until a temperature of below 1 K is reached. Starting from a random initial electron distribution, a typical annealing procedure for $n=1000$ ($N=2.43\times10^5$) in a 50$\mu$m $\times$ 50 $\mu$m box takes 20 minutes. The simulation has two input parameters: \(N\) and \(V_\text{cp}\), the second of which determines the depth of the trapping potential. We simulate several $N$'s for the range of $V_\text{cp}$ used in the experiment. Results from multiple runs with different random initial conditions were consistent. 

\begin{figure}[h!]
\centering
\includegraphics[width=0.8\textwidth]{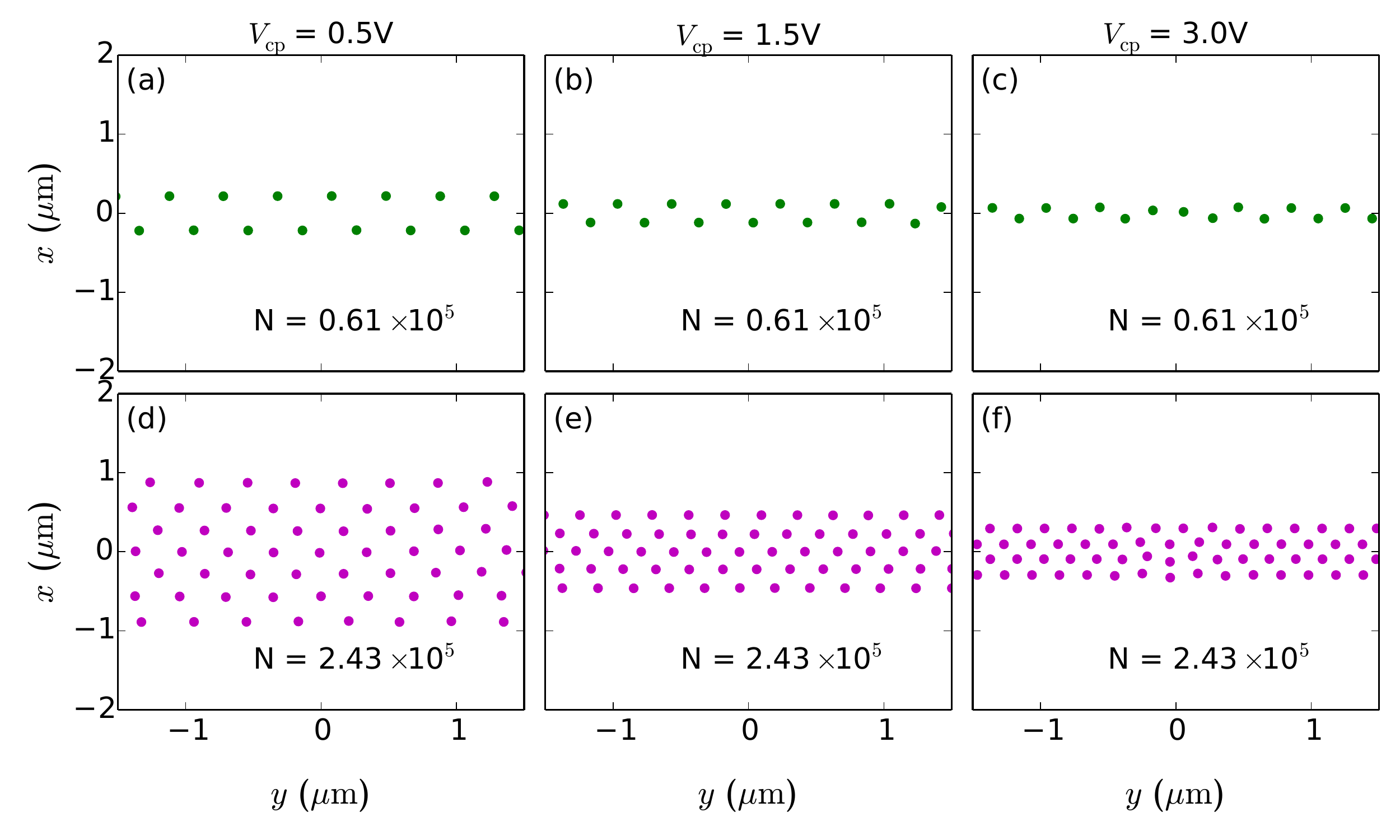}
\caption{Examples of electron configurations obtained by molecular dynamics simulations. (a)-(c) Equilibrium electron configurations for $N=0.61 \times 10^5$ electrons on the resonator as the bias voltage is increased from 0.5V, 1.5V to 3.0V, respectively. (d)-(f) Same as in (a)-(c) but for $N=2.43\times 10^5$.}
\label{fig:electron_configurations}
\end{figure}

\begin{figure}[h!]
\centering
\includegraphics[width=\textwidth]{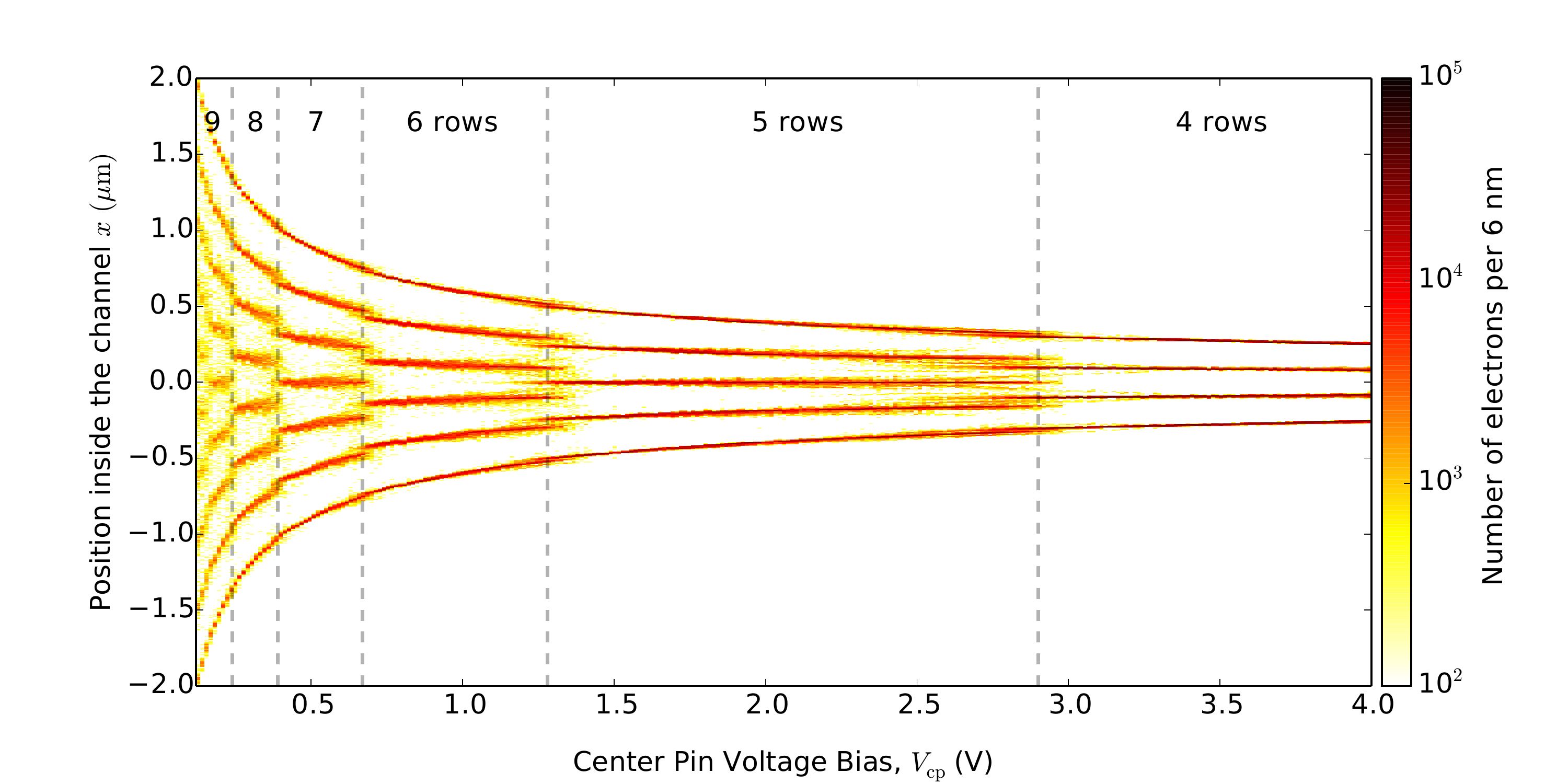}
\caption{Integrated electron distribution for a large number of electrons on the resonator: $N = 2.43\cdot10^5$. For each bias voltage the density along the $x$-axis was obtained by integrating the electron distribution along $y$. Gray dashed lines indicate transitions that lead to configurations with one fewer row. \label{fig:sample_electron_configuration}}

\end{figure}

Fig.\ref{fig:electron_configurations} shows equilibrium electron configurations for six different pairs of \{\(N\), \(V_\text{cp}\)\}. The electrons arrange themselves in rows along the $y$-direction, the direction perpendicular to the electrostatic trap. Increasing the trap bias voltage leads to an increase in electron density, a reduction in the number of rows, and a decrease of the overall width of the ensemble. This is depicted more clearly in Fig.\ref{fig:sample_electron_configuration} for $N=2.43\times10^5$. In this figure the color represents the binned electron density along $x$ (bin size 6 nm), which was obtained by integrating the electron distribution along $y$. Gradual transitions from 9 rows to 4 rows can be observed as $V_{\text{cp}}$ is swept from 0 to 4V.

\section{Equation of motion for electrons coupled to a CPW cavity}
With the electron configurations in hand, our goal is to compute the dispersive resonance 
frequency shift. One approach is to calculate all the electron normal modes, and 
use perturbation theory to find the frequency shift. This approach is difficult, as defects 
in the electron Wigner crystal and the flat potential along the $y$-direction lead to 
degenerate modes. To tackle this problem, we follow a non-perturbative approach, and calculate 
the cavity normal mode from the equations of motion of the entire electron-cavity coupled system.

\subsection{Electron Subsystem}
To calculate the equations of motion for the electrons, we use a Lagrangian formalism. The full Lagrangian $\mathcal{L}$ consists of an electronic part $\mathcal{L}_e$, a cavity part $\mathcal{L}_c$ and an interaction energy $\mathcal{L}_\text{coupling}$. The electronic part is given by
\begin{align}
    \mathcal{L}_e = \frac{1}{2} m_e \sum_i \dot{\mathbf{r}}_i^2 - e \sum_i V_{\text{DC}} (\mathbf{r}_i) - \frac{1}{2} \frac{e^2}{4\pi \epsilon_0} \sum_{i} \sum_{j\neq i}   \frac{1}{|\mathbf{r}_i-\mathbf{r}_j|},
\end{align}
where $\mathbf{r}_i = (x_i, y_i)$ is the coordinate of electron $i$ and $V_{\text{DC}} (\mathbf{r}_i)$ is the electrostatic potential that defines the static trap for the electrons. In our case $V_{\text{DC}}(\mathbf{r}) = \frac{1}{2} k_{\text{trap}} x^2$. The trap depth and thus $k_\text{trap}$ are determined by the center pin bias voltage. 

\subsection{Cavity}
Working in the charge basis, the cavity part of the Lagrangian is
\begin{align}
    \mathcal{L}_c = \frac{1}{2} L \dot{Q}^2 - \frac{Q^2}{2C},
\end{align}
where $L$ and $C$ are the effective inductance and capacitance respectively.

\subsection{Electron-Cavity Coupling}
The coupling between the electrons and the cavity can be written straightforwardly as 
\begin{align}
    \mathcal{L}_{\text{coupled}} = e \sum_i V_{\text{RF}}(\mathbf{r}_i,Q).
\end{align}
Here $V_{\text{RF}}(\mathbf{r}_i, Q)$ is the RF potential generated by the resonator. It should be emphasized that generally this potential may have a different position dependence than the electrostatic potential $V_{\text{DC}}(\mathbf{r}) = \frac{1}{2} k_\text{trap} x^2$. However, in this work the electrostatic potential is determined by the bias voltage on the center pin, whereas $V_{\text{RF}}(\mathbf{r}_i, Q)$ depends on the RF voltage (or rather charge) on the center pin. Since both potentials originate from the center pin, the functional dependence is the same.

To proceed with the analysis, the charge and position dependent parts are separated, such that
\begin{align}
    \mathcal{L}_{\text{coupled}} &= \frac{eQ}{C} \sum_i U_\text{RF}(\mathbf{r}_i).
\end{align}
$U_\text{RF}(\mathbf{r}_i)$ is a dimensionless function that describes the position dependence of the microwave potential.

\subsection{Full Equations of Motion}
The equations of motion are easily obtained by calculating the derivatives with respect to $x_i$, $y_i$, $Q$ and their time derivatives. In general
\begin{align}
    \frac{\mathrm{d}}{\mathrm{d} t}\frac{\partial \mathcal{L}}{\partial \dot{q}}-\frac{\partial\mathcal{L}}{\partial q} = 0,
\end{align}
which for the cavity results in:
\begin{align}
    L\ddot{Q} + \frac{Q}{C} + \frac{e}{C}\sum_i  U_\text{RF}(\mathbf{r}_i) = 0
    \label{eq:cavity_eom_exact}
\end{align}
In a similar fashion, the equation of motion in the $x$-direction for electron $i$ becomes
\begin{align}
    m_e \ddot{x}_i +  e \frac{\partial V_\text{DC}}{\partial x_i}(\mathbf{r}_i) + \frac{e Q}{C} \frac{\partial U_\text{RF}}{\partial x_i}(\mathbf{r}_i)  - \frac{1}{2} \frac{e^2}{4\pi \epsilon_0}  \sum_{j \neq i} \frac{x_i-x_j}{|\mathbf{r}_i-\mathbf{r}_j|^3} = 0.
    \label{eq:electrons_eom_exact}
\end{align}
Eqs. \eqref{eq:cavity_eom_exact} and \eqref{eq:electrons_eom_exact} are the exact equations of motion but are also highly nonlinear. To this end the system is linearized around the equilibrium point $(x_{i, \text{eq}}, y_{i, \text{eq}}, Q_{\text{eq}}=0)$. This means that the equations of motion will contain terms involving $\mathbf{r}_{i, \text{eq}}$. This underlines that obtaining the equilibrium electron configuration via molecular dynamics simulations is of critical importance.

At the equilibrium point the potential energy 
\begin{align}
    \mathcal{U} = \frac{Q^2}{2C} + \frac{eQ}{C} \sum_i U_\text{RF}(\mathbf{r}_i) + e\sum_i V_\text{DC} (\mathbf{r}_i) + \frac{1}{2} \frac{e^2}{4\pi \epsilon_0} \sum_{i} \sum_{j\neq i}   \frac{1}{|\mathbf{r}_i-\mathbf{r}_j|}
\end{align}
of the system is minimized, such that
\begin{align}
    &\frac{\partial \mathcal{U}}{\partial x_i} (x_{i, \text{eq}}, y_{i, \text{eq}}, Q_{\text{eq}}=0) = e \frac{\partial V_\text{DC}}{\partial x_i}( \mathbf{r}_{i, \text{eq}}) - \frac{1}{2} \frac{e^2}{4\pi \epsilon_0} \sum_{j\neq i}   \frac{x_{i, \text{eq}} - x_{j, \text{eq}} }{|\mathbf{r}_{i, \text{eq}}-\mathbf{r}_{j, \text{eq}}|^3} = 0 \label{eq:minimization_cond_xi}\\
    &\frac{\partial \mathcal{U}}{\partial y_i} (x_{i, \text{eq}}, y_{i, \text{eq}}, Q_{\text{eq}}=0) = e \frac{\partial V_\text{DC}}{\partial y_i}( \mathbf{r}_{i, \text{eq}}) - \frac{1}{2} \frac{e^2}{4\pi \epsilon_0} \sum_{j\neq i}   \frac{y_{i, \text{eq}} - y_{j, \text{eq}} }{|\mathbf{r}_{i, \text{eq}}-\mathbf{r}_{j, \text{eq}}|^3} = 0 \label{eq:minimization_cond_yi}\\
    &\frac{\partial \mathcal{U}}{\partial Q} (x_{i, \text{eq}}, y_{i, \text{eq}}, Q_{\text{eq}}=0) = \frac{e}{C} \sum_i  U_\text{RF} (\mathbf{r}_{i, \text{eq}}) = 0 \label{eq:minimization_cond_Q}
\end{align}
Finally, Eqs. \eqref{eq:cavity_eom_exact} and \eqref{eq:electrons_eom_exact} can be linearized. First consider Eq.\eqref{eq:cavity_eom_exact}. If \(\delta Q\) is a small deviation from $Q_{\text{eq}} = 0$ and $\delta x_i$ is a small deviation from $x_{i, \text{eq}}$ this equation may be written as
\begin{align}
    &L\delta \ddot{Q} + \frac{\delta Q}{C} + \frac{e}{C} \sum_i \left[ U_\text{RF} (\mathbf{r}_{i, \text{eq}}) + \frac{\partial U_\text{RF}}{\partial x_i} (\mathbf{r}_{i, \text{eq}}) \delta x_i + \frac{\partial U_\text{RF}}{\partial y_i} (\mathbf{r}_{i, \text{eq}}) \delta y_i \right] = 0, 
\end{align}
and by using Eq.\eqref{eq:minimization_cond_Q} we arrive at the equation of motion for the cavity: 
\begin{align}
    L \delta \ddot{Q} + \frac{\delta Q}{C} + \frac{e}{C} \sum_i  \left[ \frac{\partial U_\text{RF}}{\partial x_i} (\mathbf{r}_{i, \text{eq}}) \delta x_i + \frac{\partial U_\text{RF}}{\partial y_i} (\mathbf{r}_{i, \text{eq}}) \delta y_i  \right] = 0, \label{eq:eom_linearized_Q}.
\end{align}

\begin{figure}[ht]
    \centering
    \includegraphics[width=0.7\textwidth]{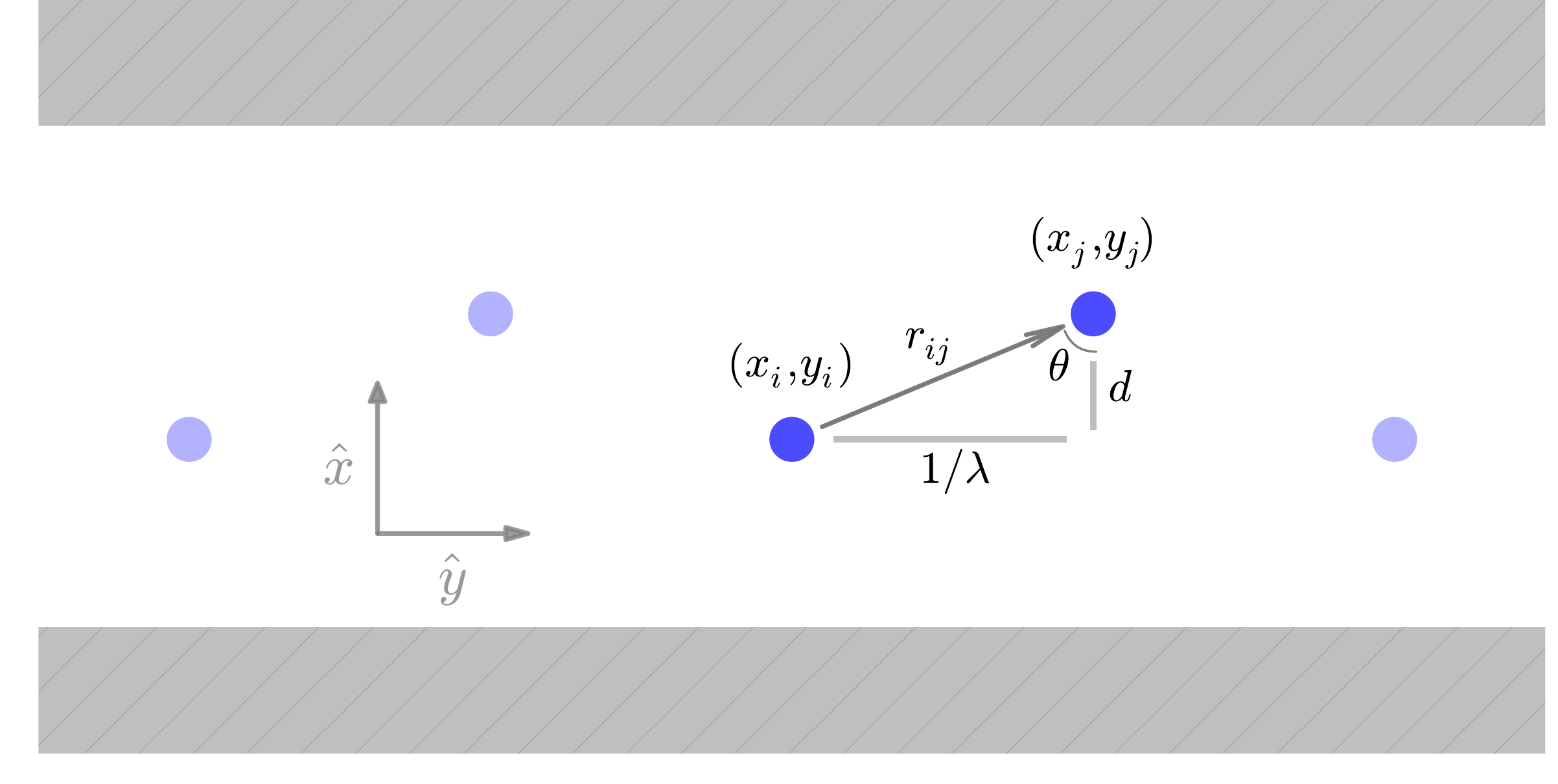}
    \caption{Schematic representation of two nearby interacting electrons located at $(x_i, y_i)$ and $(x_j, y_j)$. The distance between the two electrons is given by $|\mathbf{r}_i - \mathbf{r}_j| = r_{ij}$. The angle between the line that connects the two charges and the $x$-axis is $\theta_{ij}$. Here the trap confines electrons along the \(x\)-axis. The electron spacing along the $y$-direction is \(\frac{1}{\lambda}\) and finally the row spacing is \(d\). Note that this figure is not to scale. \label{fig:coulomb_appr_diagram}}
\end{figure}

Now for the electrons we linearize Eq.\eqref{eq:electrons_eom_exact}. To do this, it is useful to note that (see Fig.\ref{fig:coulomb_appr_diagram})
\begin{align}
    \frac{x_i-x_j}{|\mathbf{r}_i-\mathbf{r}_j|^3} &= \frac{r_{ij}\cos \theta_{ij} + (\delta x_i - \delta x_j)}{r_{ij}^3 \left[ \cos^2 \theta_{ij}\left(1 + \frac{\delta x_i - \delta x_j}{r_{ij} \cos \theta_{ij} }\right)^2 + \sin^2 \theta_{ij} \left(1 + \frac{\delta y_i - \delta y_j}{r_{ij} \sin \theta_{ij} }\right)^2 \right]^{3/2}}\notag\\ 
    &=\frac{\cos \theta_{ij}}{r_{ij}^2} - \frac{1}{2}(1+3\cos(2\theta_{ij}))\frac{\delta x_i - \delta x_j}{r_{ij}^3} - \frac{3}{2} \sin (2\theta_{ij}) \frac{\delta y_i - \delta y_j}{r_{ij}^3} +\mathcal{O}(\delta^2). \label{eq:coulomb_appr}
\end{align}
Here $r_{ij} = |\mathbf{r}_{i,\text{eq}} - \mathbf{r}_{j, \text{eq}}|$ and $\theta_{ij}$ is the angle between $\mathbf{r}_{i,\text{eq}} - \mathbf{r}_{j, \text{eq}}$ and the $x$-axis. 

The linearized version of the equation of motion then becomes
\begin{align}
     &m_e \delta\ddot{x}_i + e \left[ \frac{\partial V_\text{DC}}{\partial x_i} (\mathbf{r}_{i, \text{eq}}) +  \frac{\partial^2V_\text{DC}}{\partial x_i^2}(\mathbf{r}_{i, \text{eq}}) \delta x_i + \frac{\partial^2V_\text{DC}}{\partial x_i \partial y_i}(\mathbf{r}_{i, \text{eq}}) \delta y_i\right] + \frac{e \delta Q}{C} \frac{\partial U_\text{RF}}{\partial x_i} (\mathbf{r}_{i, \text{eq}}) \notag\\
     &- \frac{1}{2} \frac{e^2}{4\pi \epsilon_0}  \sum_{j \neq i} \frac{1}{r_{ij}^3} \left(r_{ij} \cos \theta_{ij} - \frac{1}{2} \left(1+3\cos(2\theta_{ij}) \right) \left(\delta x_i - \delta x_j\right) - \frac{3}{2} \sin (2\theta_{ij}) (\delta y_i - \delta y_j) \right) = 0
\end{align}
Using Eq.\eqref{eq:minimization_cond_xi} this can be simplified to 
\begin{align}
     &m_e \delta\ddot{x}_i+ e \frac{\partial^2 V_\text{DC}}{\partial x_i^2}(\mathbf{r}_{i, \text{eq}}) \delta x_i + e \frac{\partial^2 V_\text{DC}}{\partial x_i \partial y_i}(\mathbf{r}_{i, \text{eq}}) \delta y_i + \frac{e}{C} \frac{\partial U_\text{RF}}{\partial x_i} (\mathbf{r}_{i, \text{eq}}) \delta Q  \notag\\
     &+\frac{1}{4}\frac{e^2}{4\pi \epsilon_0}  \sum_{j \neq i} \left[\left(1+3\cos(2\theta_{ij})\right) \frac{\delta x_i-\delta x_j}{r_{ij}^3} + 3 \sin (2\theta_{ij}) \frac{\delta y_i - \delta y_j}{r_{ij}^3} \right]= 0. \label{eq:EOM_electrons_not_simplified}
\end{align}
Finally, for the sake of transparency let 
\begin{align}
    k_{ij}^\pm = \frac{1}{4} \frac{e^2}{4\pi \epsilon_0} \frac{1\pm3\cos(2\theta_{ij})}{r_{ij}^3} \quad \text{and} \quad l_{ij} = \frac{1}{4} \frac{e^2}{4\pi \epsilon_0} \frac{3\sin(2\theta_{ij})}{r_{ij}^3}
\end{align}
such that Eq.\eqref{eq:EOM_electrons_not_simplified} simplifies to the linearized equation of motion for the electrons:
\begin{align}
    &m_e \delta\ddot{x}_i + \frac{e}{C} \frac{\partial U_\text{RF}}{\partial x_i} (\mathbf{r}_{i, \text{eq}})\delta Q + \left( e\frac{\partial^2 V_\text{DC}}{\partial x_i^2}(\mathbf{r}_{i, \text{eq}}) + \sum_{j\neq i} k_{ij}^+ \right) \delta x_i   \notag \\
    &- \sum_{j \neq i} k_{ij}^+ \delta x_j + \left( e\frac{\partial^2 V_\text{DC}}{\partial x_i \partial y_i}(\mathbf{r}_{i, \text{eq}}) + \sum_{j\neq i} l_{ij} \right) \delta y_i - \sum_{j \neq i} l_{ij}\delta y_j = 0. \label{eq:eom_linearized_e}
\end{align}
Above equations show that a change in position (either in $x$ or $y$) of electron $j$ translates into an $x$ displacement of electron $i$. In the $y$-direction the equation of motion follows in a similar way: 
\begin{align}
    &m_e \delta\ddot{y}_i + \frac{e}{C} \frac{\partial U_\text{RF}}{\partial y_i} (\mathbf{r}_{i, \text{eq}})\delta Q + \left( e\frac{\partial^2 V_\text{DC}}{\partial y_i^2}(\mathbf{r}_{i, \text{eq}}) + \sum_{j\neq i} k_{ij}^- \right) \delta y_i   \notag \\
    &- \sum_{j \neq i} k_{ij}^- \delta y_j + \left(  e\frac{\partial^2 V_\text{DC}}{\partial x_i \partial y_i}(\mathbf{r}_{i, \text{eq}}) + \sum_{j\neq i} l_{ij} \right) \delta x_i - \sum_{j \neq i} l_{ij}\delta x_j = 0. \label{eq:eom_linearized_e_y}
\end{align}
The system of equations \eqref{eq:eom_linearized_Q}, \eqref{eq:eom_linearized_e} and \eqref{eq:eom_linearized_e_y} can be written in matrix form as follows: 
\begin{align}
    \mathcal{M}
    \begin{pmatrix}
         \delta \ddot{Q} \\
         \delta \ddot{x}_1 \\
         \vdots \\
         \delta \ddot{y}_1 \\
         \vdots
    \end{pmatrix}
    = -\mathcal{K} 
    \begin{pmatrix}
         \delta Q \\
         \delta x_1 \\
         \vdots \\
         \delta y_1 \\
         \vdots
    \end{pmatrix},
\end{align}
\newpage

\begin{sideways}
\begin{minipage}{\textheight}
where $\mathcal{K}$ and $\mathcal{M}$ are the kinetic and mass matrix of the system, respectively. The entries of both matrices can be easily read from the equations of motion: 
\begin{align}
\arraycolsep=0pt
    \mathcal{M} = \left(\begin{array}{ccccccc} \ora L & \ora 0 & \ora 0 & \ora \hdots & \ora 0 & \ora 0 & \ora\hdots \\
                                  \ora 0 & \gr m_e & \gr 0 & \gr \hdots & \p0 & \p0 & \p\hdots \\
                                  \ora 0 & \gr 0 & \gr m_e & \gr \hdots & \p0 & \p0 & \p\hdots \\
                                  \ora \vdots & \gr\vdots & \gr\vdots & \gr\ddots & \p\vdots & \p\vdots & \p\ddots \\
                                  \ora 0 & \p 0 & \p 0 &\p \hdots & \y m_e & \y 0 &\y \hdots \\
                                  \ora 0 & \p 0 & \p 0 & \p \hdots & \y 0 & \y m_e & \y \hdots\\
                                  \ora \vdots & \p \vdots & \p \vdots & \p \ddots & \y \vdots & \y\vdots &\y\ddots 
    \end{array} \right)
\end{align}
and
\begin{align}
\arraycolsep=0pt
    \mathcal{K} = 
    \left(\begin{array}{ccccccc} \ora \frac{1}{C} & \ora \frac{e}{C} \frac{\partial U_\text{RF}}{\partial x_1} (\mathbf{r}_{1, \text{eq}}) & \ora\frac{e}{C} \frac{\partial U_\text{RF}}{\partial x_2} (\mathbf{r}_{2, \text{eq}}) & \ora \hdots & \ora \frac{e}{C} \frac{\partial U_\text{RF}}{\partial y_1} (\mathbf{r}_{1, \text{eq}}) & \ora \frac{e}{C} \frac{\partial U_\text{RF}}{\partial y_2} (\mathbf{r}_{2, \text{eq}}) & \ora \hdots\\
    \ora \frac{e}{C} \frac{\partial U_\text{RF}}{\partial x_1} (\mathbf{r}_{1, \text{eq}}) &  \gr e\frac{\partial^2 V_\text{DC}}{\partial x_1^2}(\mathbf{r}_{1, \text{eq}}) + \sum_{j\neq 1} k_{1j}^+ & \gr -k_{12}^+ & \gr \hdots & \p e\frac{\partial^2 V_\text{DC}}{\partial x_1 \partial y_1} (\mathbf{r}_{1, \text{eq}}) + \sum_{j\neq 1} l_{1j} & \p-l_{12} & \p\hdots \\
    \ora \frac{e}{C} \frac{\partial U_\text{RF}}{\partial x_2} (\mathbf{r}_{2, \text{eq}}) & \gr-k_{21}^+ & \gr e\frac{\partial^2 V_\text{DC}}{\partial x_2^2}(\mathbf{r}_{2, \text{eq}}) + \sum_{j\neq 2} k_{2j}^+ & \gr\hdots & \p -l_{21} & \p e\frac{\partial^2 V_\text{DC}}{\partial x_2 \partial y_2} (\mathbf{r}_{2, \text{eq}}) + \sum_{j\neq 2} l_{2j} & \p \hdots\\
    \ora \vdots & \gr\vdots & \gr\vdots & \gr\ddots & \p \vdots & \p \vdots &\p \ddots \\
    \ora \frac{e}{C} \frac{\partial U_\text{RF}}{\partial y_1} (\mathbf{r}_{1, \text{eq}}) & \p e\frac{\partial^2 V_\text{DC}}{\partial x_1 \partial y_1} (\mathbf{r}_{1, \text{eq}}) + \sum_{j\neq 1} l_{1j} & \p -l_{12} & \p \hdots & \y e\frac{\partial^2 V_\text{DC}}{\partial y_1^2} (\mathbf{r}_{1, \text{eq}}) + \sum_{j\neq 1} k_{1j}^- & \y -k_{12}^- & \y \hdots \\
    \ora \frac{e}{C} \frac{\partial U_\text{RF}}{\partial y_2} (\mathbf{r}_{2, \text{eq}}) & \p -l_{21} & \p e\frac{\partial^2 V_\text{DC}}{\partial x_2 \partial y_2} (\mathbf{r}_{2, \text{eq}}) + \sum_{j\neq 2} l_{2j} & \p \hdots & \y -k_{21}^- & \y e\frac{\partial^2 V_\text{DC}}{\partial y_2^2} (\mathbf{r}_{2, \text{eq}}) + \sum_{j\neq 2} \y k_{2j}^- & \y \hdots \\
    \ora \vdots & \p\vdots & \p\vdots & \p\ddots & \y\vdots & \y\vdots & \y\ddots
    \end{array} \right).
\end{align}
Terms in $\mathcal{M}$ and $\mathcal{K}$ that represent the cavity mode and electron-cavity coupling are shaded pink. The electron $x$ and $y$ motion are shaded green and blue, respectively. Lastly, terms that involve $x-y$ and $y-x$ coupling are shaded orange. 

\end{minipage}
\end{sideways}

With the kinetic matrix and mass matrix, the eigenvalue problem 
\begin{align}
    \mathcal{M}^{-1} \mathcal{K} | \eta_i \rangle = \omega_i^2 | \eta_i \rangle \label{eq:eq_of_motion_short}
\end{align}
can be solved to obtain the normal modes $|\eta_i \rangle$ and corresponding frequencies $\omega_i$. The eigenfrequency corresponding to the mode with the highest cavity participation tells us the dispersive cavity shift.

\subsection{Constraining the Motion Along the $x$-axis}
In the previous section the most general equations of motion for a two dimensional confined electron gas were derived. In this case, the static trap is flat in the $y$-direction:
\begin{align}
    V_\text{DC}(\mathbf{r}_i)= \frac{1}{2} k_\text{trap} x_i^2,
\end{align}
and $U_\text{RF}$ has the same $\mathbf{r}$-dependence as $V_\text{DC}$:
\begin{align}
    U_\text{RF} (\mathbf{r}_i) = \beta x_i^2.
\end{align}
Since both $V_\text{DC}$ and $U_\text{RF}$ do not depend on $y_i$ the equations of motion simplify drastically. Additionally, since the trapping potential is flat along the $y$-direction and due to the presence of defects -- either intrinsic or due to imperfect annealing -- one can get modes with imaginary or zero frequency. To reduce the impact of these modes we ``freeze'' the modes in the $y$-direction, i.e. we set $\delta y_i = 0$ in what follows to simplify the equations of motion even further. 

After these simplifications the matrices that govern the equations of motion read
\begin{align}
    \mathcal{M} = \begin{pmatrix} L & 0 & 0 & \hdots \\
                                  0 & m_e & 0 & \hdots \\
                                  0 & 0 & m_e & \hdots \\
                                  \vdots & \vdots & \vdots & \ddots
    \end{pmatrix}
\end{align}
and
\begin{align}
    \mathcal{K} = 
    \begin{pmatrix} 
        \frac{1}{C} & \frac{2 e \beta x_{1, \text{eq}}}{C}  & \frac{2 e \beta x_{2, \text{eq}}}{C} & \frac{2 e \beta x_{3, \text{eq}}}{C}& \hdots \\
        \frac{2 e \beta x_{1, \text{eq}}}{C} &  ek_\text{trap} + \sum_{j\neq 1} k_{1j} & -k_{12} & -k_{13} & \hdots \\
        \frac{2 e \beta x_{2, \text{eq}}}{C} & -k_{21} & ek_\text{trap} + \sum_{j\neq 2} k_{2j} & -k_{23} & \hdots \\
        \frac{2 e \beta x_{3, \text{eq}}}{C} & -k_{31} & -k_{32} &ek_\text{trap} + \sum_{j\neq 3} k_{3j} & \hdots \\
                                      \vdots & \vdots & \vdots & \vdots & \ddots
    \end{pmatrix}.
\end{align}

\section{Analytical Model of Simple Electron Configurations}
With only two rows of electrons inside the channel, it is possible to find the lowest energy configuration analytically. This serves as a good comparison for our numerically simulated electron configurations. Let the transverse direction of the trap be \(x\), then the trap is parabolic along the \(x\) direction and flat along \(y\). If we assume no electron loss and only two rows, as we vary the trap bias, only the transverse configuration of the electrons changes. The electron density along the \(y\) axis (\(\lambda\)) remains constant.

In the simple geometry depicted in Fig.\ref{fig:coulomb_appr_diagram}, the force electron $i$ exerts on its neighbor $j$ along \(x\) is
\begin{align}
F_{x, \,\text{single neighbor}} = \frac{1}{4 \pi \epsilon_0} \frac{e^2}{r_{ij}^2} \frac{d}{r_{ij}}.
\end{align}
Since electron $j$ has another neighbor to the right, the resulting force is close to \(2F_{x, ee}\). If we take into account of the next-nearest neighbor, the total force one electron experiences is
\begin{align}
F_{x, \text{total}} = \gamma F_{x, \,\text{single neighbor}},
\end{align}
where \(\gamma \approx 2.08\). Depending on the electron density, \(\gamma\) grows from 2.08 in the sparse limit to around 2.38 in the dense limit where the ensemble transits into 3 rows. 

\begin{figure}[h]
\centering
\includegraphics[width=0.6\textwidth]{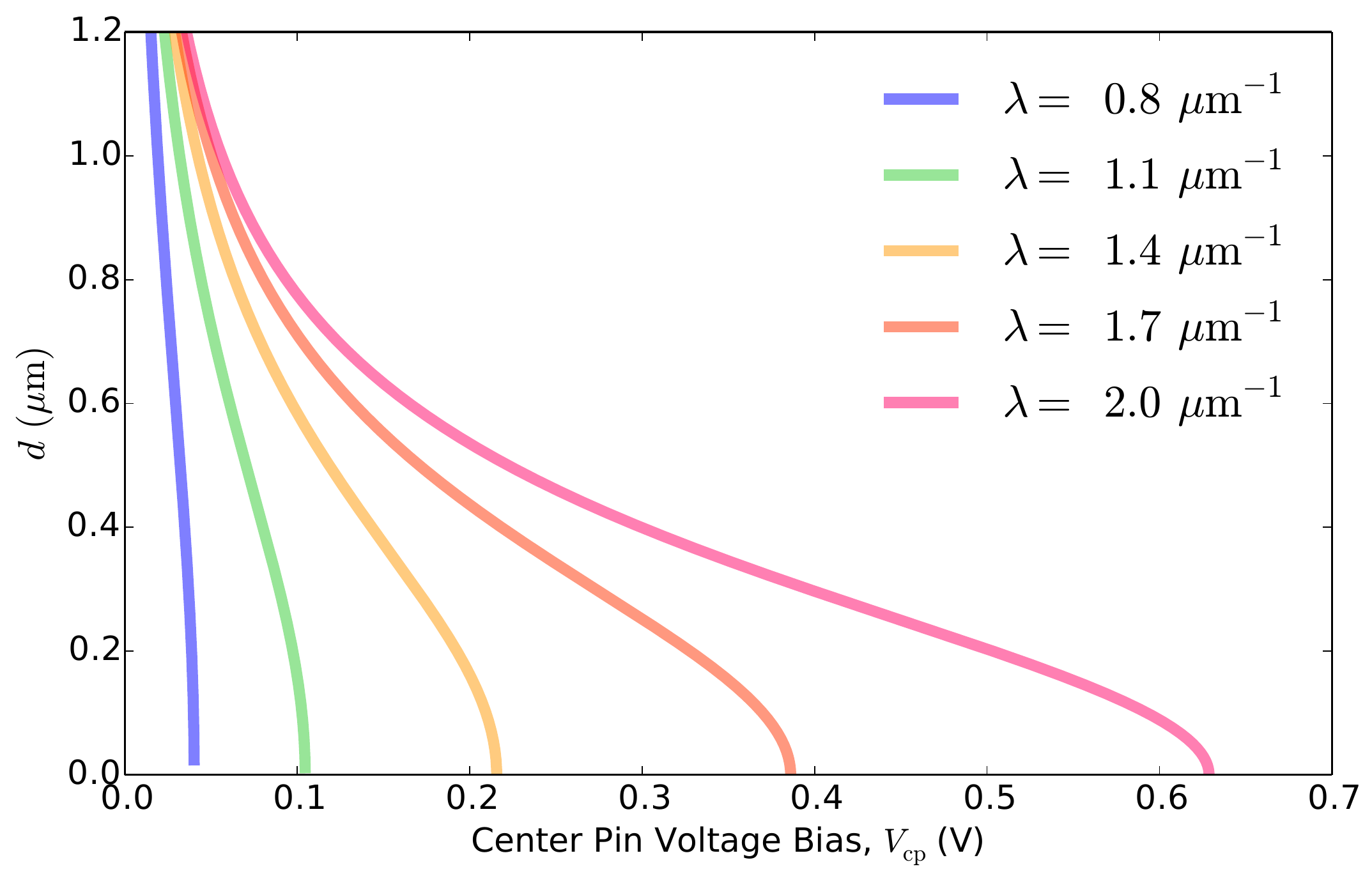}
\caption{2-row ensemble width as a function of trap bias voltage \(V_{\text{cp}}\) at various electron densities. An electron density of $\lambda = $0.8 (2.0) $\mu\mathrm{m}^{-1}$ corresponds to $N = 9.7 \times 10^3$ ($N=2.4\times10^4$) electrons on the resonator. \label{fig:width_v_res_v}}
\end{figure}

Now to find the equilibrium configuration, we have 
\begin{align}
- F_{x,\text{trap}}=\gamma F_{x, \,\text{single neighbor}} \label{eq:analytical_force_balance}.
\end{align}
Note that the \(x\)-location of the electrons is measured from the center line, therefore \(x=d/2.\) We can then rewrite Eq.\eqref{eq:analytical_force_balance} as
\begin{align}
e k_{\text{trap}} \frac{d}{2}  =\frac{\gamma e^2}{4 \pi \epsilon_0 } \frac{1}{(1/\lambda)^2 + d^2} \frac{d}{r_{ij}}.
\end{align}
Here \(k_{\text{trap}}\) is the curvature of the trap, which grows linearly with respect to the trapping bias voltage \(V_{\text{cp}}\). For any bias voltage, we can now solve for \(d\) in the equilibrium configuration
\begin{align}
d^2 = \left(\frac{\gamma}{k_\text{trap}} \frac{e}{2\pi \epsilon_0}\right)^{2/3} - (1/\lambda)^2. \label{eq:ensemble_width}
\end{align}
In Fig. \ref{fig:width_v_res_v} we plot the width of the ensemble for different densities $\lambda$. 

For simple configurations that have only \(1 \sim 2\)  rows, we derived the analytical solution of the electrostatic equilibrium configuration. However, as the electron density and the number of rows increase, the total number of interactive terms that need to be taken into account quickly grows to become unmanageable. More importantly, because the trapping potential is more shallow on the side than in the middle, adjacent rows of the electron ensemble are not commensurate. This means when there are more than 2 rows, the electrons cannot form a perfect crystal in a channel with parabolic transverse profile and our analytical solution breaks down. Therefore, for more than 2 rows we calculate the equilibrium position of the electrons numerically using our molecular dynamics formalism. 

\section{Modeling Electron-Induced Cavity frequency Shift}
\begin{figure}[h]
\centering
\includegraphics[width=\textwidth]{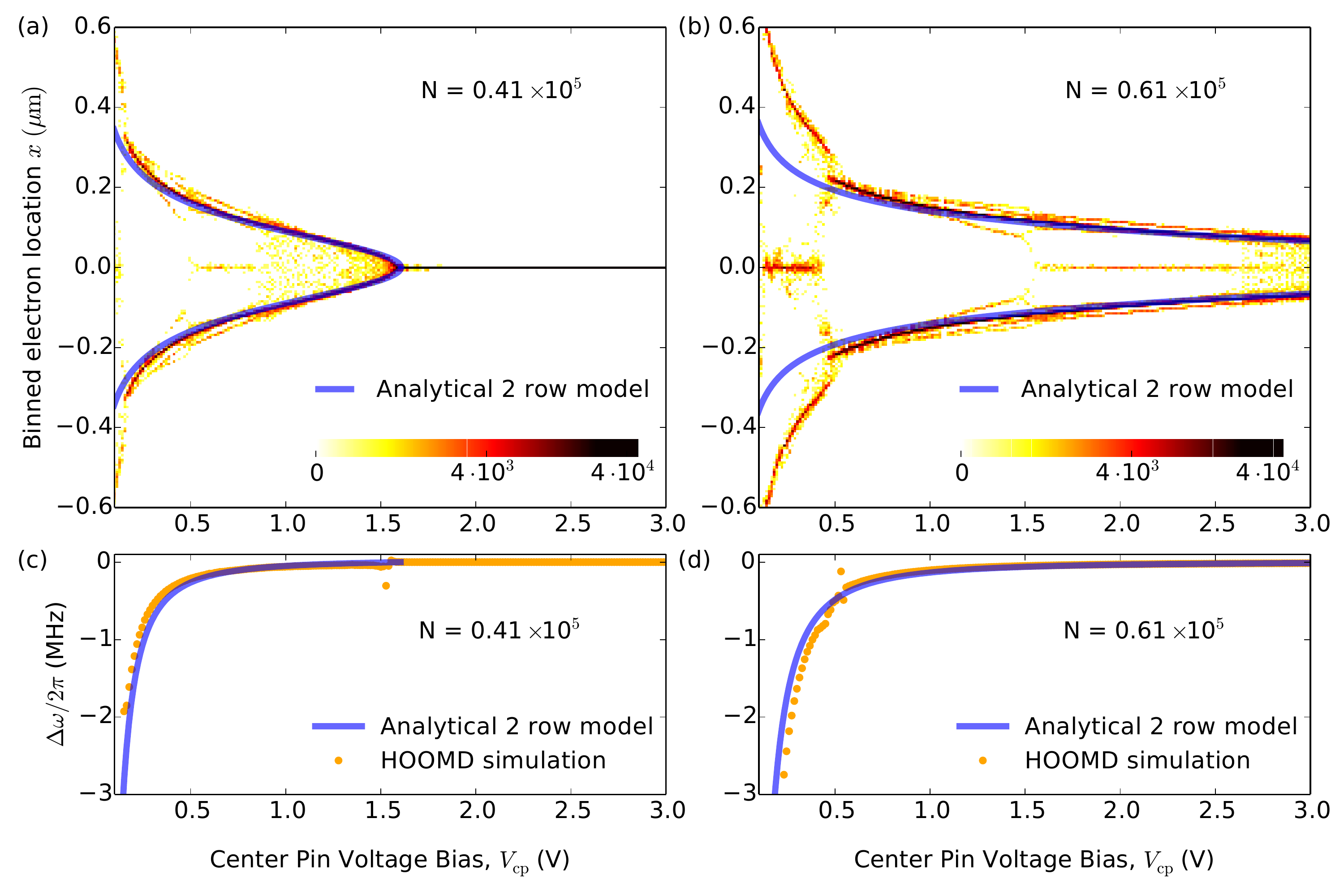}
\caption{Comparison between the analytical model and the molecular dynamics simulation for two different numbers of trapped electrons: $N=0.41\cdot10^5$ (left column) and $N=0.61\cdot10^5$ (right column). Top: binned electron density (bin size = 6 nm) as a function of $x$ via \textsc{hoomd} simulation (color in the background) and calculated width of a two row electron configuration obtained from the analytic model (blue). Note that there is no free parameter. Bottom: Cavity frequency shift solved numerically (with \textsc{hoomd}, orange dots) and analytically (blue). In the analytical solution, a simplified version of the equations of motion was used. \label{fig:hoomd_analytical_comparison}}
\end{figure}

Using the equation of motion of the constrained electron-cavity coupled system, we can now calculate the electron-induced cavity frequency shift at various bias voltages \(V_{\text{cp}}\) with different number of electrons \(N\). First, we calculate the normal mode frequency of a simple 2-row electron ensemble, using the analytical solution to the equilibrium configuration (Eq.\eqref{eq:ensemble_width}). Then we compare this result with the normal mode frequency calculated from electron configurations obtained using \textsc{hoomd}. The results are depicted in Fig.\ref{fig:hoomd_analytical_comparison}. 

In Fig.\ref{fig:hoomd_analytical_comparison}a and \ref{fig:hoomd_analytical_comparison}b., we compare the \(x\)-locations of the electrons in the ensemble from \textsc{hoomd} with those from the analytic model. In the analytical model, we only take into account nearest neighbor interactions, such that the geometric parameter $\gamma$ in Eq.\eqref{eq:ensemble_width} is equal to 2.08. Clearly, this results in good agreement in the region where $V_\text{cp} > 1.0$V. For lower bias voltages it is energetically more favorable to form three rows instead.

Next we compare the cavity frequency shift for both cases. The cavity frequency shift is obtained by solving the equation of motion, Eq.\eqref{eq:eq_of_motion_short}. For the analytical model we exploit symmetry to simplify the kinetic matrix $\mathcal{K}$. Therefore, the problem reduces to diagonalizing a 2 x 2 matrix $\mathcal{M}^{-1} \mathcal{K}$, where
\begin{align}
    \mathcal{K} = \begin{pmatrix} \omega_0^2 L & d e \beta \omega_0^2 L \sqrt{N}\\ d e \beta \omega_0^2 L \sqrt{N}&ek_\text{trap} + 2 k_{12} \end{pmatrix}.
\end{align}
Here $\omega_0 = \sqrt{1/LC}$ is the resonance frequency of the resonator and $\beta = 0.0733$ m$^{-2}$ from Eq.\eqref{eq:static_potential_function}.

The results are shown in Figures \ref{fig:hoomd_analytical_comparison}c and \ref{fig:hoomd_analytical_comparison}d. Again, both plots show good agreement between the analytical model and molecular dynamics simulation. The largest deviation occurs when the bias voltage is low, where the electron configuration transits from 3 to 2 rows. As expected, the 2-row analytic model no longer describes the correct geometry in this regime.

\subsection{Comparison with experimental data at higher densities}
The electron ensembles encountered in our experiment are much more complicated than the simple cases mentioned above. However, using the molecular dynamics simulation package HOOMD \cite{HOOMD}, we are able to anneal systems far beyond the simple 2-row case, and solve for the cavity frequency shift using these configurations. 

\begin{figure}[h]
\centering
\includegraphics[width=0.7\textwidth]{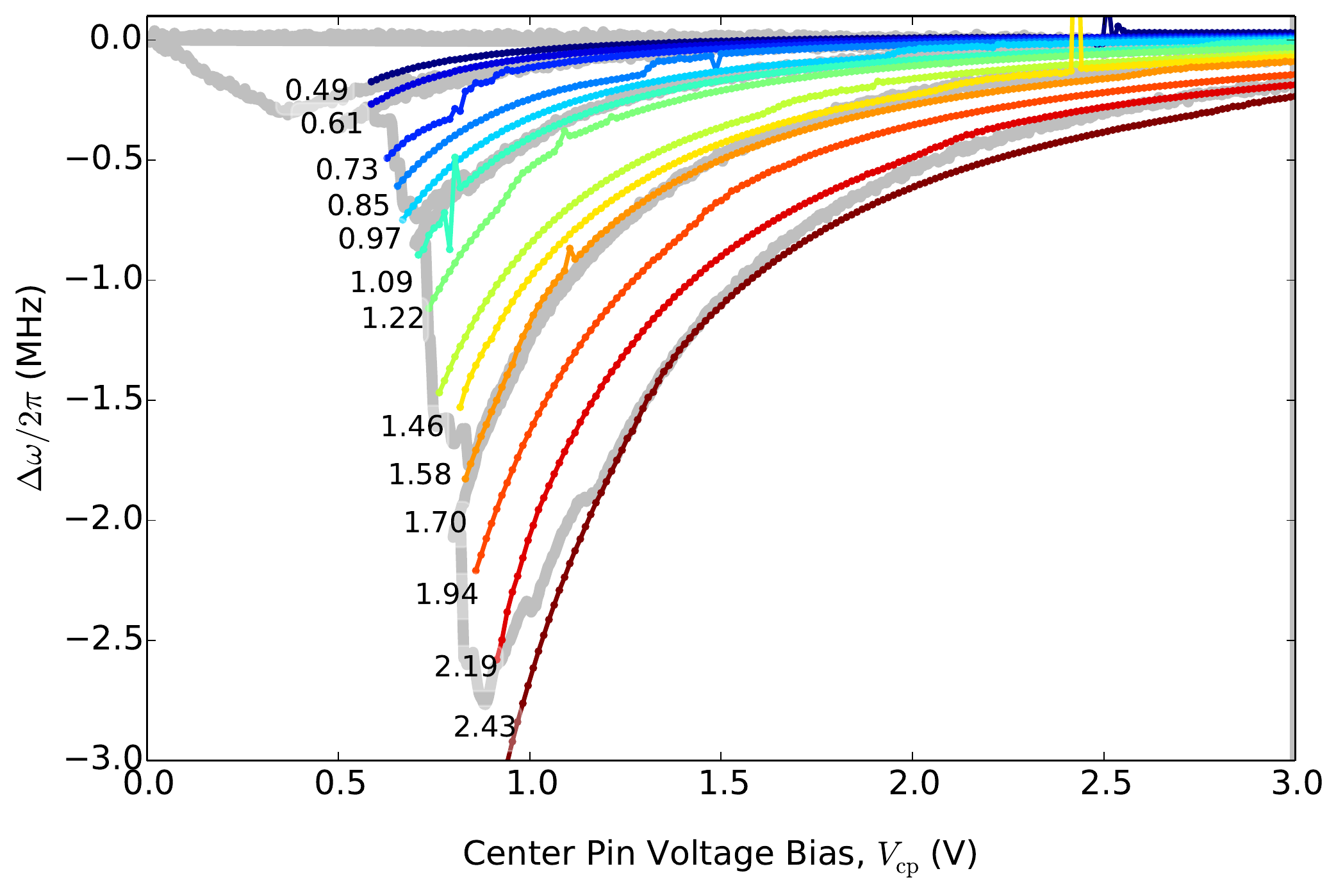}
\caption{Comparison of the measured cavity shift (gray) to the simulated cavity shift (color). The number of electrons in the simulation was increased from $N=0.49\times10^5$ (blue curve) to $N=2.43\times 10^5$. For the intermediate curves the number of electrons $N$ (in units of $10^5$) is depicted next to each curve. \label{fig:experiment_with_all_simulations}}
\end{figure}

The key observations in our experiment are the smooth change of the cavity frequency shift during the bias voltage sweep, and the sharp jumps corresponding to irreversible electron loss from the trap. In our numerical model, we are able to replicate both of these phenomena by calculating the cavity frequency shift for each point in our \(V_\text{cp}\) sweep, and for various number of electrons \(N\) in the trap. For each curve in Fig. \ref{fig:experiment_with_all_simulations}, we do not rescale the $x$ or $y$ axis nor is there any offset applied. The only free parameter is \(N\). By comparing the simulation with the experimental data, we are able to estimate how many electrons are present for each trace in the experiment shown in the main text. The four curves that match the experimental data have $N=0.61 \times 10^5$, $N=1.09 \times 10^5$, $N=1.70 \times 10^5$ and $N = 2.43 \times 10^5$.

\begin{figure}[h!]
\centering
\includegraphics[width=0.55\textwidth]{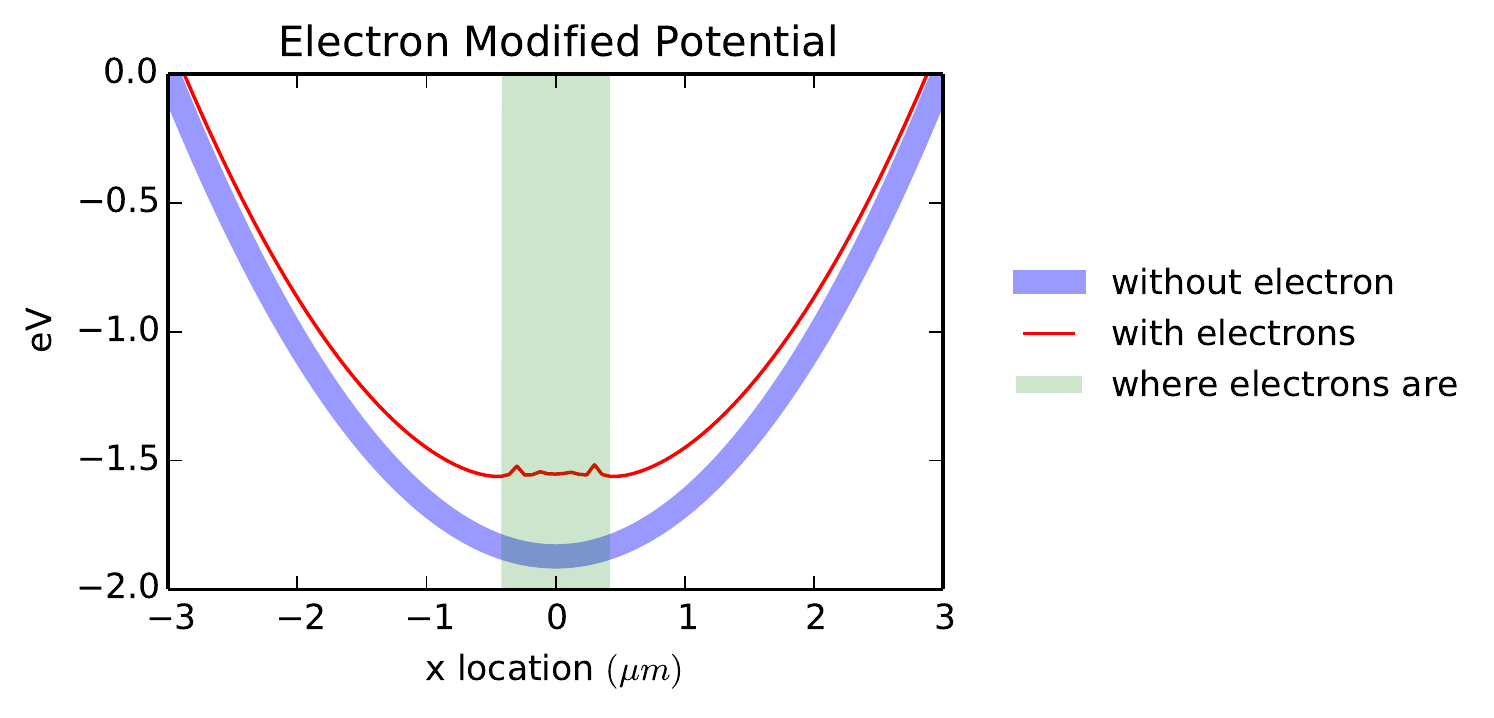}
\caption{Comparision between the trap potential (light blue) and the screened potential when electrons are present (red). A green rectangle in the middle shows where the electrons appear. The rugged shape of the screened potential is due to the discreteness of the electron configuration. \label{fig:modified_trap_potential}}
\end{figure}

In the experiment, an important observation is the loss of electrons from the trap. With the equilibrium positions, such a process can be simulated. Let us assume there is a leak in the trap with a threshold voltage \(V_\text{leak}\) that is independent of the trap bias. If the screened potential (Fig. \ref{fig:modified_trap_potential}) that an electron feels is higher than this threshold voltage, it is energetically more favorable for this electron to leak out of the trap. Hence the configuration is unstable. To determine the threshold voltage, we compute the screened potential for all electrons in each configuration. Then we find the threshold voltage that produces the best fit to the experimental data.

\begin{figure}[h!]
\centering
\includegraphics[width=0.4\textwidth]{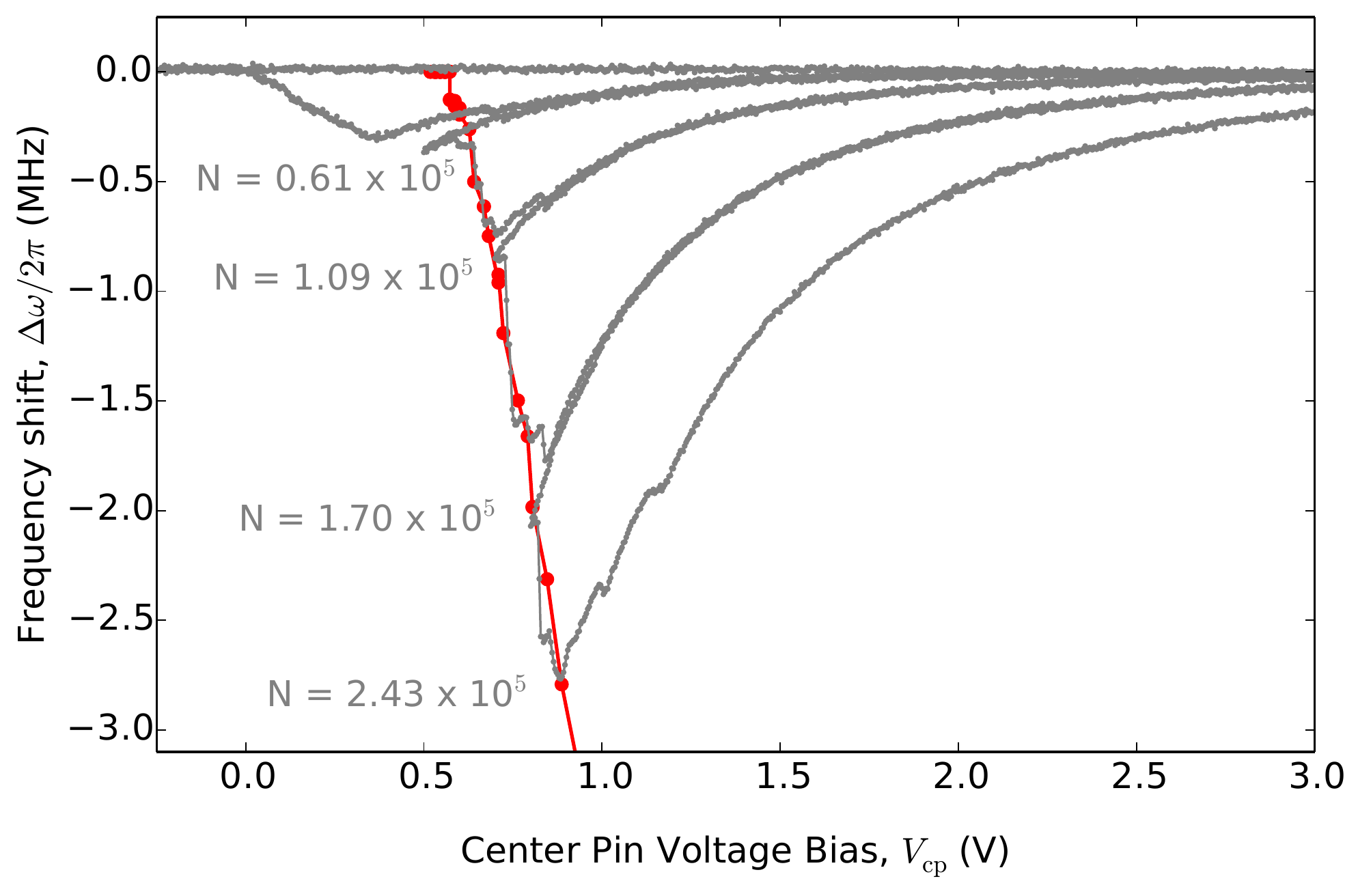}
\caption{Loss frontier calculated using \(V_{\text{leak}} = 530\) mV. The model fits well with the hysteresis data until the electron ensemble is reduced down to mostly a single row in the trap. \label{fig:hysteresis_with_loss_only}}
\end{figure}

The actual value of the leak voltage determines the starting point of each of the curves in Fig. \ref{fig:experiment_with_all_simulations}. A leak voltage of 530 mV gives the best fit to the data, shown in Fig. \ref{fig:hysteresis_with_loss_only}. 

\subsection{Normal modes of the electron configuration and coupling strength}

The same normal mode solution to the electron-cavity coupled system also gives access to the relevant modes in the electron subsystem. By picking the ten most strongly coupled electron modes, we can estimate the electron normal mode frequency during the voltage sweep.

\begin{figure}[h]
\centering
\includegraphics[width=1\textwidth]{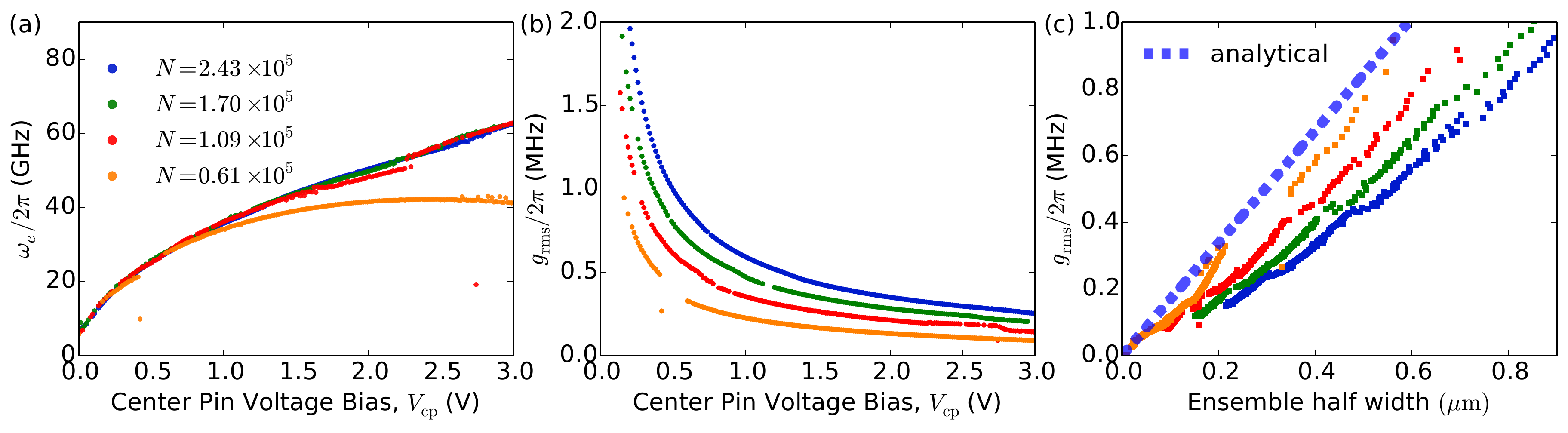}
\caption{(a): Electron normal mode frequency for $N=0.61\times 10^5$ (orange), $N=1.09\times 10^5$ (red), $N=1.70\times10^5$ (green) and $N=2.43\times 10^5$ (blue). (b): rms coupling per electron for different values of $N$. Same color coding as in (a). (c): rms coupling per electron plotted as a function of the electron configuration width. Same color coding as in (a). Small gaps in the solution appear for $V_\text{cp}$ that correspond to row transitions. For these points the unstable equilibrium configuration results in solutions where a small fraction of the electrons has a disproportionate displacement. \label{fig:electron_normal_mode_and_coupling}}
\end{figure}

In Fig.\ref{fig:electron_normal_mode_and_coupling}a we plot the normal mode frequency of the trapped electrons. For most $V_{\text{cp}}$ the electron mode frequency is tens of GHz, indicating that we are working in the dispersive limit. In this limit the cavity shift is given by $N g_{\text{rms}}^2/\Delta$, where $g_{\text{rms}}$ is the rms electron-cavity coupling and $\Delta$ is the frequency difference between the electron normal mode and the cavity resonance. A closer look at the electron mode evolution for $N=0.61\times 10^5$ reveals an interesting feature. For this curve, the electron density is low enough to support only 1 or 2 rows of electrons. As we increase the bias voltage, the electron normal mode initially grows as $\sqrt{V_{\text{cp}}}$, but rolls off as the angle \(\theta\) between each electron and its nearest neighbor approaches \(\pi/2\) (see Fig.\ref{fig:coulomb_appr_diagram}). At $\theta = \pi/2$ the electrons form a single row such that the coupling to the cavity vanishes. The voltage at which this happens is determined by the electron density \(\lambda\) and can be modeled precisely with our analytical model.

For higher $N$, where there are more than 2 electron rows, the electron normal modes evolve mostly as $\sqrt{V_{\text{cp}}}$. Minor row-reconfigurations occur, which result in small jumps in the normal mode frequency. 

Using the fact that we are in the dispersive limit, we can now look at the rms coupling per electron at different \(V_\text{cp}\) and different \(N\). This is depicted in Fig.\ref{fig:electron_normal_mode_and_coupling}b. In general, the coupling per electron decreases as the bias voltage increases and higher \(N\) leads to a higher overall coupling. Upon further analysis, we found that this trend is mostly due to the linear relationship between the location of an individual electron and its coupling to the cavity. In our normal mode solution, for an electron located at a equilibrium position \(x_{i,\text{eq}}\), the coupling term between the electron and a single photon excitation in the cavity is 
\[
\frac{2 e \beta x_{i, \text{eq}}}{C}.
\]
If we plot the rms coupling per electron as a function of the overall ensemble width, the relationship is roughly linear for constant $N$, while ensembles with more rows have a smaller rms coupling.

\section{Experimental Setup}
The measurements are done dispersively by measuring the shift in the cavity resonance frequency.

An Agilent network analyzer is connected to the input and output ports of the cavity with attenuation along the input line for photon thermalization and amplification on the output. Low-pass filters filled with the Eccosorb epoxy are used to filter out thermal photons that would otherwise enter the sample box and affect the cavity Q.

\begin{figure}[h]
\centering
    \includegraphics[width=0.50\textwidth]{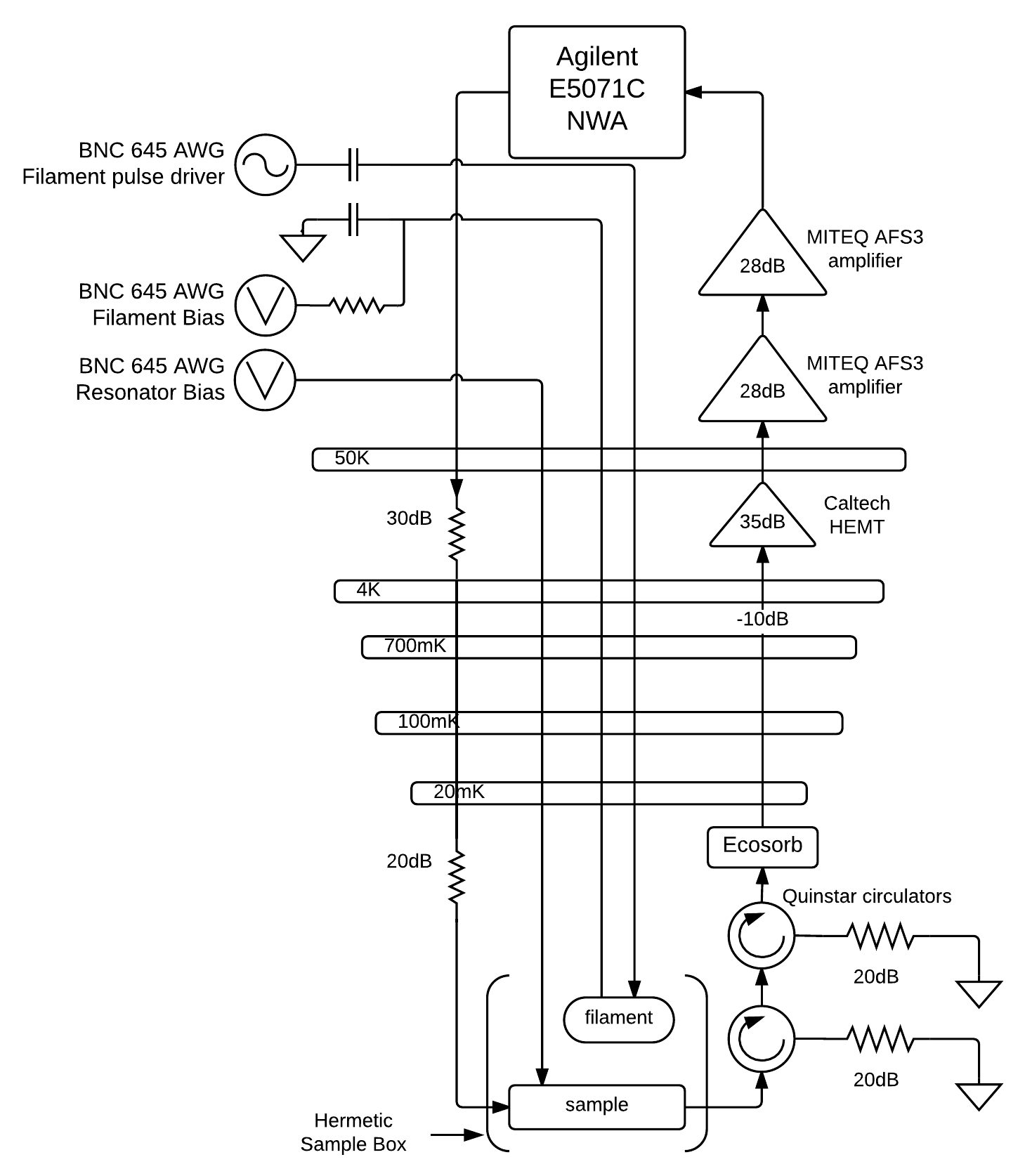}
    \raisebox{10mm}{
        \includegraphics[width=0.45\textwidth]{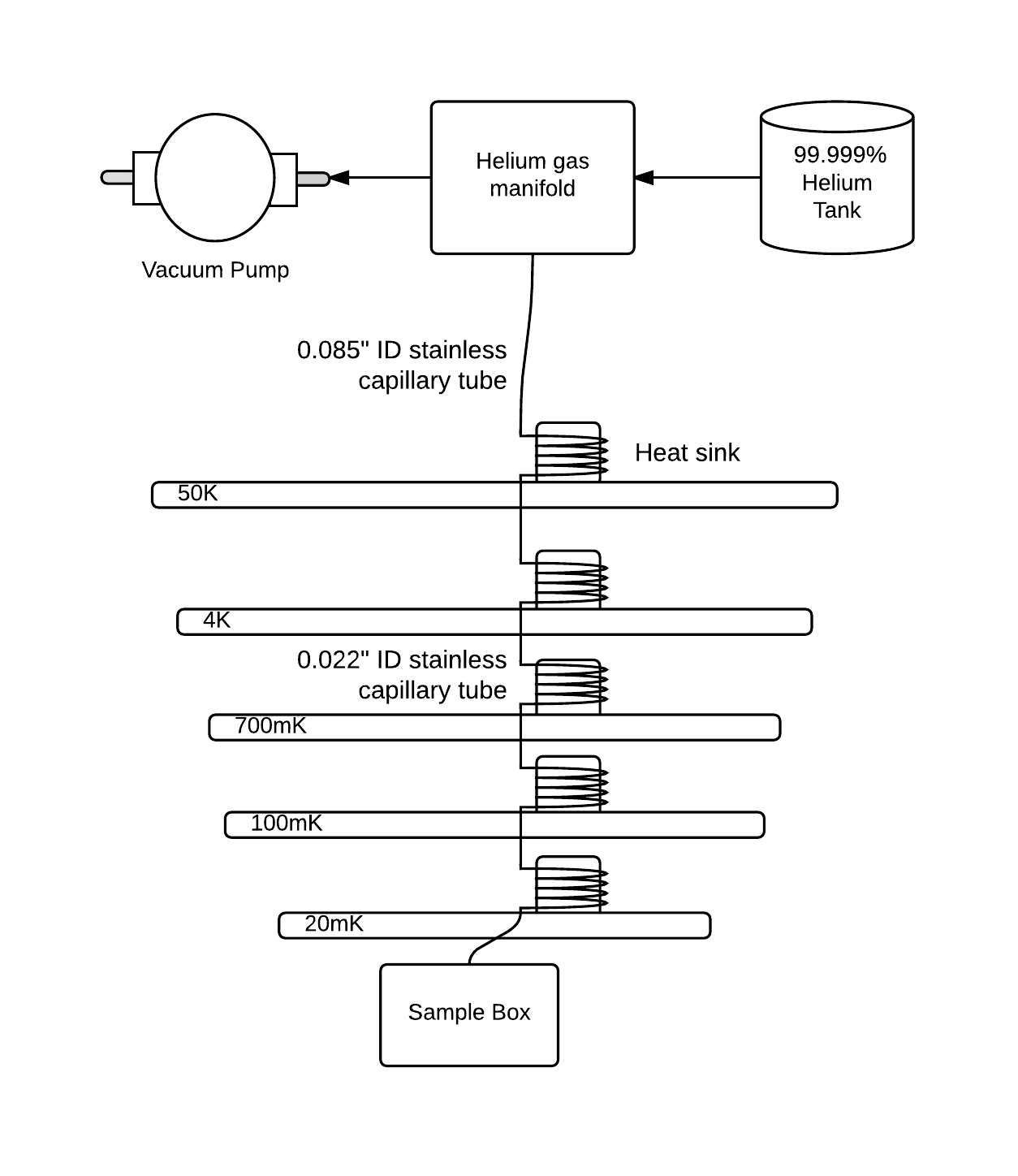}
    }
    \caption{Left: microwave measurement setup, right: gas handling system.}
\end{figure}

To feed helium into the hermetic sample box, particular consideration was taken to thermalize the room temperature helium gas with each stage of the fridge. Stainless steel capillary tube was wrapped around and soldered to copper  cylinders at each stage of the fridge. Above 4K, 0.085" ID (1/8" OD) tubes were used to prevent plugs, while below 4K the capillary tube had an ID of 0.022" (1/16" OD).

During the experiment, the transmission spectrum of the cavity is monitored using a network analyzer (NWA). The electrons are loaded by pulsing a small filament briefly with a negative bias voltage. As the sample cools down after the loading, we record the cavity resonance frequency and the quality factor.

\section{Device Fabrication}
The coplanar waveguide resonator chip is fabricated via a two-step e-beam lithography process. The ground plane is made of \(800\,\text{nm}\) thick niobium, whereas the center pin of the waveguide is made of \(80 \,\text{nm}\) aluminum. All patterning is done on a JOEL \(100 \,\text{keV}\) e-beam writer. 

Due to the thickness of the niobium layer \((800 \,\text{nm})\), an aluminum hard mask is required for the etching. Re-depositioning of aluminum (micro-masking) occurred during the RIE process, which is mitigated by over-etching at the end.

Below we list all the fabrication steps:

\begin{enumerate}
\item First Layer
\begin{enumerate}
\item Nb deposition and wafer preparation
\item Coat 2-inch sapphire wafer with 800 nm Nb
\item Coat wafer with  aluminum as the dry etch stopper layer
\end{enumerate}
\item E-beam lithography
\begin{enumerate}
\item Spin coat ZEP 520 at 3000 rpm for 45 s
\item Bake resist at 150C for 180 s
\item Pattern with a JEOL JBX9300FS tool; area dose 1600 C/cm\(^2\).
\item Develop ZEP in xylene at \(0^{\circ}\text{C}\), rinse with DI water.
\item Oxygen plasma clean 10 min
\end{enumerate}
\item Dry etch (reactive ion etching)
\begin{enumerate}
\item Prepare the chamber by running the Al etch recipe for 10 min.
\item To pattern the aluminum hard mask use \(\text{BCl}_3\)(3.0 sccm) and \(\text{Cl}_2\)(24.0 sccm) at 20$^\circ$C, for 1:45 min.
\item Prepare the chamber by running the Nb etch recipe for 10 min.
\item Then with \(\text{SF}_6\)(25 sccm) + Ar (5 sccm) at 20$^\circ$C, for 15 min until etched through.
\end{enumerate}
\item Second layer:
\begin{enumerate}
\item E-beam lithography
\begin{enumerate}
\item Spin coat ZEP 520 at 4000 rpm for 45 s.
\item Pattern with JOEL tool at area dose of 1600 C/cm\(^2\).
\item Develop ZEP at \(0^{\circ}\text{C}\), rinse with IPA and DI water.
\item Oxygen plasma clean 10 min
\end{enumerate}
\item Aluminum lift-off
\begin{enumerate}
\item Evaporate 80 nm of aluminum
\item Dip wafer in n-methylpyrrolidone (NMP) heated to \(60 ^{\circ}\text{C}\),
\item As the resist dissolves away, gently blow away the aluminum and dip into acetone heated to \(60^{\circ}\text{C}\)
\item take out and blow dry.
\end{enumerate}
\end{enumerate}

\end{enumerate}

\bibliographystyle{naturemag} 
\bibliography{main.bib}

\providecommand{\noopsort}[1]{}\providecommand{\singleletter}[1]{#1}%
\begin{thebibliography}{10}
\expandafter\ifx\csname url\endcsname\relax
  \def\url#1{\texttt{#1}}\fi
\expandafter\ifx\csname urlprefix\endcsname\relax\def\urlprefix{URL }\fi
\providecommand{\bibinfo}[2]{#2}
\providecommand{\eprint}[2][]{\url{#2}}

\bibitem{Lyon2006}
\bibinfo{author}{Lyon, S.~A.}
\newblock \bibinfo{title}{{Spin-based quantum computing using electrons on
  liquid helium}}.
\newblock \emph{\bibinfo{journal}{Physical Review A}}
  \textbf{\bibinfo{volume}{74}}, \bibinfo{pages}{1--6} (\bibinfo{year}{2006}).
\newblock \urlprefix\url{http://link.aps.org/doi/10.1103/PhysRevA.74.052338
  http://dx.doi.org/10.1103/PhysRevA.74.052338}.

\bibitem{Platzman1999}
\bibinfo{author}{Platzman, P.~M.} \& \bibinfo{author}{Dykman, M.~I.}
\newblock \bibinfo{title}{{Quantum Computing with Electrons Floating on Liquid
  Helium}}.
\newblock \emph{\bibinfo{journal}{Science}} \textbf{\bibinfo{volume}{284}},
  \bibinfo{pages}{1967--1969} (\bibinfo{year}{1999}).
\newblock \urlprefix\url{http://www.ncbi.nlm.nih.gov/pubmed/10373109
  http://www.sciencemag.org/content/284/5422/1967}.

\bibitem{Dykman2003}
\bibinfo{author}{Dykman, M.~I.}, \bibinfo{author}{Platzman, P.~M.} \&
  \bibinfo{author}{Seddighrad, P.}
\newblock \bibinfo{title}{{Qubits with electrons on liquid helium}}.
\newblock \emph{\bibinfo{journal}{Physical Review B}}
  \textbf{\bibinfo{volume}{67}}, \bibinfo{pages}{155402}
  (\bibinfo{year}{2003}).
\newblock \urlprefix\url{http://link.aps.org/doi/10.1103/PhysRevB.67.155402}.

\bibitem{Schuster2010b}
\bibinfo{author}{Schuster, D.~I.}, \bibinfo{author}{Fragner, A.},
  \bibinfo{author}{Dykman, M.~I.}, \bibinfo{author}{Lyon, S.~A.} \&
  \bibinfo{author}{Schoelkopf, R.~J.}
\newblock \bibinfo{title}{{Proposal for Manipulating and Detecting Spin and
  Orbital States of Trapped Electrons on Helium Using Cavity Quantum
  Electrodynamics}}.
\newblock \emph{\bibinfo{journal}{Physical Review Letters}}
  \textbf{\bibinfo{volume}{105}}, \bibinfo{pages}{1--4} (\bibinfo{year}{2010}).
\newblock
  \urlprefix\url{http://link.aps.org/doi/10.1103/PhysRevLett.105.040503}.

\bibitem{Monarkha2004}
\bibinfo{author}{Monarkha, Y.} \& \bibinfo{author}{Kono, K.}
\newblock \emph{\bibinfo{title}{{Two-Dimensional Coulomb Liquids and Solids}}}
  (\bibinfo{publisher}{Springer}, \bibinfo{address}{Berlin},
  \bibinfo{year}{2004}).
\newblock \urlprefix\url{http://www.springer.com/us/book/9783540207542}.

\bibitem{Shirahama1995a}
\bibinfo{author}{Shirahama, K.} \& \bibinfo{author}{Kono, K.}
\newblock \bibinfo{title}{{Dynamical Transition in the Wigner Solid on a Liquid
  Helium Surface}}.
\newblock \emph{\bibinfo{journal}{Physical Review Letters}}
  \textbf{\bibinfo{volume}{74}}, \bibinfo{pages}{781--784}
  (\bibinfo{year}{1995}).
\newblock \urlprefix\url{http://link.aps.org/doi/10.1103/PhysRevLett.74.781}.

\bibitem{Shirahama1995b}
\bibinfo{author}{Shirahama, K.}, \bibinfo{author}{Ito, S.},
  \bibinfo{author}{Suto, H.} \& \bibinfo{author}{Kono, K.}
\newblock \bibinfo{title}{{Surface study of liquid3He using surface state
  electrons}}.
\newblock \emph{\bibinfo{journal}{Journal of Low Temperature Physics}}
  \textbf{\bibinfo{volume}{101}}, \bibinfo{pages}{439--444}
  (\bibinfo{year}{1995}).
\newblock \urlprefix\url{http://link.springer.com/10.1007/BF00753334}.

\bibitem{williams1971}
\bibinfo{author}{Williams, R.}, \bibinfo{author}{Crandall, R.} \&
  \bibinfo{author}{Willis, A.}
\newblock \bibinfo{title}{{Surface states of electrons on liquid helium}}.
\newblock \emph{\bibinfo{journal}{Physical Review Letters}}
  (\bibinfo{year}{1971}).
\newblock \urlprefix\url{http://link.aps.org/doi/10.1103/PhysRevLett.26.7}.

\bibitem{Grimes1979}
\bibinfo{author}{Grimes, C.~C.} \& \bibinfo{author}{Adams, G.}
\newblock \bibinfo{title}{{Evidence for a Liquid-to-Crystal Phase Transition in
  a Classical, Two-Dimensional Sheet of Electrons}}.
\newblock \emph{\bibinfo{journal}{Physical Review Letters}}
  \textbf{\bibinfo{volume}{42}}, \bibinfo{pages}{795--798}
  (\bibinfo{year}{1979}).
\newblock \urlprefix\url{http://link.aps.org/doi/10.1103/PhysRevLett.42.795}.

\bibitem{Andrei1997}
\bibinfo{author}{Andrei}.
\newblock \emph{\bibinfo{title}{{Two-Dimensional Electron Systems}}},
  vol.~\bibinfo{volume}{19} (\bibinfo{year}{1997}).
\newblock \urlprefix\url{http://link.springer.com/10.1007/978-94-015-1286-2}.

\bibitem{dwyer2013}
\bibinfo{author}{Ikegami, H.}, \bibinfo{author}{Tsutsumi, Y.} \&
  \bibinfo{author}{Kono, K.}
\newblock \bibinfo{title}{{Chiral Symmetry Breaking in Superfluid 3He-A}}.
\newblock \emph{\bibinfo{journal}{Science}} \textbf{\bibinfo{volume}{341}},
  \bibinfo{pages}{59--62} (\bibinfo{year}{2013}).
\newblock
  \urlprefix\url{http://www.sciencemag.org/cgi/doi/10.1126/science.1236509}.

\bibitem{kono2010}
\bibinfo{author}{Kono, K.}
\newblock \bibinfo{title}{{Electrons on the Surface of Superfluid 3He}}.
\newblock \emph{\bibinfo{journal}{Journal of Low Temperature Physics}}
  \textbf{\bibinfo{volume}{158}}, \bibinfo{pages}{288--300}
  (\bibinfo{year}{2010}).
\newblock \urlprefix\url{http://link.springer.com/10.1007/s10909-009-9962-3}.

\bibitem{chepelianskii2015}
\bibinfo{author}{Chepelianskii, A.~D.}, \bibinfo{author}{Watanabe, M.},
  \bibinfo{author}{Nasyedkin, K.}, \bibinfo{author}{Kono, K.} \&
  \bibinfo{author}{Konstantinov, D.}
\newblock \bibinfo{title}{{An incompressible state of a photo-excited electron
  gas}}.
\newblock \emph{\bibinfo{journal}{Nature Communications}}
  \textbf{\bibinfo{volume}{6}}, \bibinfo{pages}{7210} (\bibinfo{year}{2015}).
\newblock \urlprefix\url{http://www.nature.com/doifinder/10.1038/ncomms8210}.

\bibitem{Papageorgiou2005}
\bibinfo{author}{Papageorgiou, G.} \emph{et~al.}
\newblock \bibinfo{title}{{Counting individual trapped electrons on liquid
  helium}}.
\newblock \emph{\bibinfo{journal}{Applied Physics Letters}}
  \textbf{\bibinfo{volume}{86}}, \bibinfo{pages}{1--3} (\bibinfo{year}{2005}).

\bibitem{Rousseau2009}
\bibinfo{author}{Rousseau, E.} \emph{et~al.}
\newblock \bibinfo{title}{{Addition spectra of Wigner islands of electrons on
  superfluid helium}}.
\newblock \emph{\bibinfo{journal}{Physical Review B}}
  \textbf{\bibinfo{volume}{79}}, \bibinfo{pages}{045406}
  (\bibinfo{year}{2009}).
\newblock \urlprefix\url{http://link.aps.org/doi/10.1103/PhysRevB.79.045406}.

\bibitem{Bradbury2011}
\bibinfo{author}{Bradbury, F.~R.} \emph{et~al.}
\newblock \bibinfo{title}{{Efficient Clocked Electron Transfer on Superfluid
  Helium}}.
\newblock \emph{\bibinfo{journal}{Physical Review Letters}}
  \textbf{\bibinfo{volume}{107}}, \bibinfo{pages}{266803}
  (\bibinfo{year}{2011}).
\newblock
  \urlprefix\url{http://link.aps.org/doi/10.1103/PhysRevLett.107.266803}.
\newblock \eprint{1107.4040}.

\bibitem{Takita2014}
\bibinfo{author}{Takita, M.} \& \bibinfo{author}{Lyon, S.~a.}
\newblock \bibinfo{title}{{Isolating electrons on superfluid helium}}.
\newblock \emph{\bibinfo{journal}{Journal of Physics: Conference Series}}
  \textbf{\bibinfo{volume}{568}}, \bibinfo{pages}{052034}
  (\bibinfo{year}{2014}).
\newblock
  \urlprefix\url{http://stacks.iop.org/1742-6596/568/i=5/a=052034?key=crossref.2fec45ff72d23fff62247d21b092b69b}.

\bibitem{Gorkov1973}
\bibinfo{author}{Gor'kov, L.} \& \bibinfo{author}{Chernikova, D.}
\newblock \bibinfo{title}{{Concerning the structure of a charged surface of
  liquid helium}}.
\newblock \emph{\bibinfo{journal}{Soviet Journal of Experimental \ldots}}
  (\bibinfo{year}{1973}).
\newblock \urlprefix\url{http://adsabs.harvard.edu/abs/1973JETPL..18...68G}.

\bibitem{Edelman1980}
\bibinfo{author}{Edel'man, V.~S.}
\newblock \bibinfo{title}{{Levitated electrons}}.
\newblock \emph{\bibinfo{journal}{Soviet Physics Uspekhi}}
  \textbf{\bibinfo{volume}{23}}, \bibinfo{pages}{227--244}
  (\bibinfo{year}{1980}).
\newblock
  \urlprefix\url{http://stacks.iop.org/0038-5670/23/i=4/a=R01?key=crossref.248a1a88e99c7bc7faeb6c8f819ebe25}.

\bibitem{Petta2005a}
\bibinfo{author}{Petta, J.~R.}
\newblock \bibinfo{title}{{Coherent Manipulation of Coupled Electron Spins in
  Semiconductor Quantum Dots}}.
\newblock \emph{\bibinfo{journal}{Science}} \textbf{\bibinfo{volume}{309}},
  \bibinfo{pages}{2180--2184} (\bibinfo{year}{2005}).
\newblock \urlprefix\url{http://www.ncbi.nlm.nih.gov/pubmed/16141370
  http://www.sciencemag.org/cgi/doi/10.1126/science.1116955}.

\bibitem{Wallraff2004}
\bibinfo{author}{Wallraff, A.} \emph{et~al.}
\newblock \bibinfo{title}{{Strong coupling of a single photon to a
  superconducting qubit using circuit quantum electrodynamics}}.
\newblock \emph{\bibinfo{journal}{Nature}} \textbf{\bibinfo{volume}{431}},
  \bibinfo{pages}{162--167} (\bibinfo{year}{2004}).
\newblock \urlprefix\url{http://dx.doi.org/10.1038/nature02851}.

\bibitem{Blais2004}
\bibinfo{author}{Blais, A.}, \bibinfo{author}{Huang, R.-S.},
  \bibinfo{author}{Wallraff, A.}, \bibinfo{author}{Girvin, S.~M.} \&
  \bibinfo{author}{Schoelkopf, R.~J.}
\newblock \bibinfo{title}{Cavity quantum electrodynamics for superconducting
  electrical circuits: An architecture for quantum computation}.
\newblock \emph{\bibinfo{journal}{Phys. Rev. A}} \textbf{\bibinfo{volume}{69}},
  \bibinfo{pages}{062320} (\bibinfo{year}{2004}).
\newblock \urlprefix\url{http://link.aps.org/doi/10.1103/PhysRevA.69.062320}.

\bibitem{Sommer1971}
\bibinfo{author}{Sommer, W.} \& \bibinfo{author}{Tanner, D.}
\newblock \bibinfo{title}{{Mobility of Electrons on the Surface of Liquid
  \^{}\{4\}He}}.
\newblock \emph{\bibinfo{journal}{Physical Review Letters}}
  \textbf{\bibinfo{volume}{27}}, \bibinfo{pages}{1345--1349}
  (\bibinfo{year}{1971}).
\newblock \urlprefix\url{http://link.aps.org/doi/10.1103/PhysRevLett.27.1345}.

\bibitem{Schoelkopf2008}
\bibinfo{author}{Schoelkopf, R.~J.} \& \bibinfo{author}{Girvin, S.~M.}
\newblock \bibinfo{title}{{Wiring up quantum systems}}.
\newblock \emph{\bibinfo{journal}{Nature}} \textbf{\bibinfo{volume}{451}},
  \bibinfo{pages}{664--669} (\bibinfo{year}{2008}).
\newblock \urlprefix\url{http://www.ncbi.nlm.nih.gov/pubmed/18256662
  http://dx.doi.org/10.1038/451664a}.

\bibitem{Devoret2013}
\bibinfo{author}{Devoret, M.~H.} \& \bibinfo{author}{Schoelkopf, R.~J.}
\newblock \bibinfo{title}{Superconducting circuits for quantum information: An
  outlook}.
\newblock \emph{\bibinfo{journal}{Science}} \textbf{\bibinfo{volume}{339}},
  \bibinfo{pages}{1169--1174} (\bibinfo{year}{2013}).
\newblock
  \urlprefix\url{http://www.sciencemag.org/content/339/6124/1169.abstract}.
\newblock \eprint{http://www.sciencemag.org/content/339/6124/1169.full.pdf}.

\bibitem{Marty1986}
\bibinfo{author}{Marty, D.}
\newblock \bibinfo{title}{{Stability of two-dimensional electrons on a
  fractionated helium surface}}.
\newblock \emph{\bibinfo{journal}{Journal of Physics C: Solid State Physics}}
  \textbf{\bibinfo{volume}{19}}, \bibinfo{pages}{6097--6104}
  (\bibinfo{year}{1986}).
\newblock
  \urlprefix\url{http://stacks.iop.org/0022-3719/19/i=30/a=019?key=crossref.c8fb6fb583aac58a288a9a74e9c3fafc}.

\bibitem{Rees2011}
\bibinfo{author}{Rees, D.} \emph{et~al.}
\newblock \bibinfo{title}{{Point-Contact Transport Properties of Strongly
  Correlated Electrons on Liquid Helium}}.
\newblock \emph{\bibinfo{journal}{Physical Review Letters}}
  \textbf{\bibinfo{volume}{106}}, \bibinfo{pages}{1--4} (\bibinfo{year}{2011}).
\newblock
  \urlprefix\url{http://link.aps.org/doi/10.1103/PhysRevLett.106.026803}.

\bibitem{Armour2013}
\bibinfo{author}{Armour, a.~D.}, \bibinfo{author}{Blencowe, M.~P.},
  \bibinfo{author}{Brahimi, E.} \& \bibinfo{author}{Rimberg, a.~J.}
\newblock \bibinfo{title}{{Universal Quantum Fluctuations of a Cavity Mode
  Driven by a Josephson Junction}}.
\newblock \emph{\bibinfo{journal}{Physical Review Letters}}
  \textbf{\bibinfo{volume}{111}}, \bibinfo{pages}{247001}
  (\bibinfo{year}{2013}).
\newblock
  \urlprefix\url{http://link.aps.org/doi/10.1103/PhysRevLett.111.247001}.
\newblock \eprint{1307.2498}.

\bibitem{Petersson2012}
\bibinfo{author}{Petersson, K.~D.} \emph{et~al.}
\newblock \bibinfo{title}{{Circuit quantum electrodynamics with a spin qubit}}.
\newblock \emph{\bibinfo{journal}{Nature}} \textbf{\bibinfo{volume}{490}},
  \bibinfo{pages}{380--383} (\bibinfo{year}{2012}).
\newblock \urlprefix\url{http://dx.doi.org/10.1038/nature11559}.

\bibitem{Anderson2008}
\bibinfo{author}{Anderson, J.~a.}, \bibinfo{author}{Lorenz, C.~D.} \&
  \bibinfo{author}{Travesset, A.}
\newblock \bibinfo{title}{{General purpose molecular dynamics simulations fully
  implemented on graphics processing units}}.
\newblock \emph{\bibinfo{journal}{Journal of Computational Physics}}
  \textbf{\bibinfo{volume}{227}}, \bibinfo{pages}{5342--5359}
  (\bibinfo{year}{2008}).
\newblock
  \urlprefix\url{http://linkinghub.elsevier.com/retrieve/pii/S0021999108000818}.

\bibitem{HOOMD}
\bibinfo{title}{{HOOMD}}.
\newblock \urlprefix\url{https://codeblue.umich.edu/hoomd-blue/}.

\end{thebibliography}

\end{document}